\documentclass[12pt]{article}
\usepackage{amsmath}%\pdfoutput=1
\usepackage{cite}
\begin{document}
\begin{flushright}
KANAZAWA-15-16\\
December, 2015
\end{flushright}
\vspace*{1cm}
  
\begin{center} 
{\Large\bf Radiative neutrino mass model with degenerate right-handed neutrinos}
\vspace*{1cm}

{\Large Shoichi Kashiwase}\footnote{e-mail:~shoichi@hep.s.kanazawa-u.ac.jp}
{\Large and Daijiro Suematsu}\footnote{e-mail:~suematsu@hep.s.kanazawa-u.ac.jp}
\vspace*{1cm}\\

{\it Institute for Theoretical Physics, Kanazawa University, 
\\ Kanazawa 920-1192, Japan}
\end{center}
\vspace*{1.5cm} 

\noindent
{\Large\bf Abstract}\\
The radiative neutrino mass model can relate neutrino masses and 
dark matter at a TeV scale. If we apply this model to thermal leptogenesis,
we need to consider resonant leptogenesis at that scale.
It requires both finely degenerate masses for the right-handed neutrinos and 
a tiny neutrino Yukawa coupling. We propose an extension of the model 
with a U(1) gauge symmetry, in which these conditions are shown to be 
simultaneously realized through a TeV scale symmetry breaking. 
Moreover, this extension can bring about a small quartic scalar coupling 
between the Higgs doublet scalar and an inert doublet scalar which 
characterizes the radiative neutrino mass generation. It also is 
the origin of the $Z_2$ symmetry which guarantees the stability of dark matter. 
Several assumptions which are independently supposed in the original model 
are closely connected through this extension. 

\newpage
%%%%%%%%%%%%%%%%%%%%%%%%%%%%%%%%%%%
\section{Introduction}
ATLAS and CMS groups in the LHC experiment have reported the 
discovery of the Higgs-like particle \cite{lhc}. All the 
standard model contents seem to have been found by now. 
However, the standard model 
has serious problems from experimental and observational view points.
Although the existence of neutrino masses and dark matter has been 
confirmed through various experiments and observations 
\cite{nexp,t13,uobs,planck}, 
it cannot be explained in the standard model. The standard model 
cannot give a framework for the generation of baryon number 
asymmetry in the Universe, either \cite{basym}. 
These facts now cause serious tension between the standard model and Nature 
so that they motivate us to consider its extension.  

The radiative neutrino mass model proposed in \cite{ma} is a simple and 
interesting extension of the standard model which could be an explanation.
In several previous articles \cite{ndm,u1,susyndm,ndm1,ks,infl}, 
we have studied these problems in this model and its extensions.
They suggest that these problems could be explained in a consistent way, 
simultaneously. 
Unfortunately, however, we could not justify
several assumptions and the parameter tuning adopted in these explanations. 
For example, if we consider thermal leptogenesis in this model, both finely
degenerate right-handed neutrino masses and a small Yukawa coupling for
the lightest right-handed neutrino are required in order to 
make possible sufficient generation of
lepton number asymmetry through the out-of-equilibrium decay 
of the lightest right-handed neutrino. 
In this work, we have just assumed them
independently in a way consistent with other phenomenological issues.

In this paper, we consider an extension of the model which makes it
possible to realize these required conditions simultaneously in the 
evolution of the Universe. 
We suppose a new symmetry breaking at a scale of $O(1)$ TeV for this purpose.
After this symmetry breaking, a small mass difference is induced between
two lighter right-handed neutrinos, although they have an equal mass 
originally. At the same time, a Yukawa coupling of the lightest right-handed 
neutrino becomes much smaller than that of the heavier one. 
To realize this scenario, we introduce a low energy U(1) 
gauge symmetry to the model. 
We show that (i) both the almost degenerate right-handed neutrino masses 
and a tiny neutrino Yukawa coupling, which are indispensable 
for TeV scale resonant leptogenesis \cite{res},
are brought about after the breaking of this symmetry. 
Moreover, we find that this extension can also explain important key 
features required in the original Ma model, 
that is, (ii) a small quartic coupling between the Higgs doublet scalar 
and an inert doublet scalar which plays a crucial role in the neutrino mass
generation, and (iii) the origin of the $Z_2$ symmetry which guarantees 
the stability of dark matter. 

The remaining part of this paper is organized as follows. 
After introducing an extended model
in the next section, we discuss features in the scalar sector and also 
the right-handed neutrino mass degeneracy.
Baryon number asymmetry generated through the thermal leptogenesis is studied 
taking account of these. 
In section 3, we study the dark matter relic abundance and other cosmological 
aspects of the model. Finally, in section 4 we give a brief summary of
the main results of the paper.

\section{An extended model}
\subsection{U(1) gauge symmetry at a TeV scale}
The original Ma model is a simple extension of the standard model
which can relate neutrino masses and dark matter \cite{ma}.
In this model, only an inert doublet scalar $\eta$ and right-handed 
neutrinos $N_i$ are added to the standard model.
Although ingredients of the standard model are assigned an even parity of
the imposed $Z_2$ symmetry, new fields are assumed to have odd parity.
This feature forbids tree-level neutrino mass generation and
guarantees the stability of dark matter.
 
We extend this model with a U(1)$_X$ gauge symmetry, a singlet scalar
$S$, and also additional right-handed neutrinos $\tilde N_i$ whose
number is equal to the one of $N_i$. 
The U(1)$_X$ charge is assigned each new ingredient as $Q_X(S)=2$, 
$Q_X(\eta)=-1$, $Q_X(N_i)=1$, and $Q_X(\tilde N_i)=-1$.
Normalization for the U(1)$_X$ charge and coupling is fixed 
through a covariant derivative, which is defined as
$D_\mu=\partial_\mu-ig\frac{\tau^a}{2}W_\mu^a-ig_Y\frac{Y}{2}B_\mu
-ig_X\frac{Q_X}{2}X_\mu$.
Since the standard model fields are 
assumed to have no charge for this U(1)$_X$, it is obvious that 
the U(1)$_X$ is anomaly free.
If this symmetry is assumed to break down due to a vacuum expectation value
$\langle S\rangle$,
the model has a remnant exact symmetry $Z_2$ after this breaking. 
Since only $\eta$, $N_i$ and $\tilde N_i$
have its odd parity, the lightest one of them is stable and can be dark matter.
We assume that dark matter is the lightest neutral component of $\eta$ 
in this study.

The relevant part of the Lagrangian for these new ingredients of the model is 
summarized as
\begin{eqnarray}
-{\cal L}_N&=&h_{\alpha i} \bar{N_i}\eta^\dagger\ell_{\alpha}
+f_{\alpha i}\frac{S^\dagger}{M_\ast}\bar{\tilde {N_i}}
\eta^\dagger \ell_{\alpha}+M_iN_i\tilde N_i
+\frac{y_i}{2}S^\dagger N_iN_i + \frac{\tilde y_i}{2}S\tilde{N_i}\tilde{N_i}
+ {\rm h.c.}, \nonumber \\
V&=&\lambda_1(\phi^\dagger\phi)^2+\lambda_2(\eta^\dagger\eta)^2
+\lambda_3(\phi^\dagger\phi)(\eta^\dagger\eta) 
+\lambda_4(\eta^\dagger\phi)(\phi^\dagger\eta)  
+\frac{\lambda_5^\prime}{2}
\Big[\frac{S}{M_\ast}(\phi^\dagger\eta)^2 + {\rm h.c.}\Big]
\nonumber \\
&+&\lambda_6(S^\dagger S)(\phi^\dagger\phi) + \lambda_7(S^\dagger S)
(\eta^\dagger\eta) +\kappa(S^\dagger S)^2
+m_\phi^2\phi^\dagger\phi +m_\eta^2\eta^\dagger\eta +m_S^2S^\dagger S,
\label{model}
\end{eqnarray}
where $\ell_{\alpha}$ is a left-handed doublet lepton  
and $\phi$ is an ordinary doublet Higgs scalar.
$M_\ast$ is a cut-off scale of this model.
The bare masses $M_i$ and $m_\eta$ 
in eq.~(\ref{model}) are assumed to be real and of $O(1)$~TeV.
The couplings $h_{\alpha i}$ and $f_{\alpha i}$ in the neutrino sector 
are considered to be written by using the basis in which 
the Yukawa coupling matrix of charged leptons is diagonal. 
As easily found in eq.~(\ref{model}), if the singlet $S$ has a vacuum
expectation value, the coupling $\lambda_5$ in the original Ma model 
and neutrino Yukawa couplings $\tilde h_{\alpha i}$ for $\tilde N_i$ 
are determined as \cite{u1} 
\begin{equation}
\lambda_5=\lambda_5^\prime\frac{\langle S\rangle}{M_\ast}, \qquad
\tilde h_{\alpha i}=f_{\alpha i}\frac{\langle S^\dagger\rangle}{M_\ast},
\label{l5c}
\end{equation}
where it may be natural to consider that both $\lambda_5^\prime$ and 
$f_{\alpha i}$ are of $O(1)$.   
The magnitude of $\lambda_5$ is crucial for the neutrino mass determination 
in the model. We note that it can be small enough if 
$|\langle S\rangle|\ll M_\ast$ is satisfied. 
Scales assumed for $|\langle S\rangle|$ and $M_\ast$ in the present study 
are discussed below. 

\subsection{Scalar sector}  
First, we discuss the scalar sector of the model.
We express the scalar fields by using a unitary gauge 
\begin{equation}
\phi^T=(0, \langle\phi\rangle +\frac{h}{\sqrt{2}}), \quad 
\eta^T=(\eta^+, \frac{1}{\sqrt 2}(\eta_R+i\eta_I)), \quad
S=\langle S\rangle + \frac{\sigma}{\sqrt{2}}, 
\label{unitary}
\end{equation}
where both vacuum expectation values $\langle\phi\rangle$ 
and $\langle S\rangle$ are assumed to be real and positive.
In this vacuum, the new Abelian gauge boson $X_\mu$ gets a mass
$m_X^2=2g_X^2\langle S\rangle^2$.
The scalar potential $V$ in eq.~(\ref{model}) can be represented by
using eq.~(\ref{unitary}) as
\begin{eqnarray}
V&=& \frac{1}{2}\left( 4\lambda_1\langle\phi\rangle^2h^2+
4\kappa\langle S\rangle^2\sigma^2 
+ 4\lambda_6\langle\phi\rangle\langle S\rangle h\sigma\right)
+\frac{1}{2}M_{\eta_R}^2\eta_R^2+ \frac{1}{2}M_{\eta_I}^2\eta_I^2
+ M_{\eta_c}^2\eta^+\eta^- \nonumber \\
&+&\frac{1}{4}\left[\sqrt{\lambda_1}h^2-\sqrt{\lambda_2}
(2\eta^+\eta^-+\eta_R^2+\eta_I^2)-\sqrt\kappa\sigma^2\right]^2 
+\frac{1}{4}\left[\left\{2(\lambda_3+2\sqrt{\lambda_1\lambda_2})\eta^+\eta^-
\right.\right. \nonumber \\
&+&\left.\left.(\lambda_++2\sqrt{\lambda_1\lambda_2})\eta_R^2+
(\lambda_-+2\sqrt{\lambda_1\lambda_2})\eta_I^2
+(\lambda_6+2\sqrt{\lambda_1\kappa})\sigma^2\right\}h^2\right. \nonumber \\
&+&\left.(\lambda_7-2\sqrt{\lambda_2\kappa})(2\eta_+\eta_-+\eta_R^2
+\eta_I^2)\sigma^2\right] 
+\sqrt{2}\lambda_1\langle\phi\rangle h^3
+\sqrt 2\kappa\langle S\rangle \sigma^3 \nonumber \\
&+&{\sqrt 2}(\lambda_3 \langle\phi\rangle h + 
\lambda_7\langle S\rangle\sigma)\eta_+\eta_-  
+\frac{1}{\sqrt 2}\left(\lambda_+\langle\phi\rangle h 
+\lambda_7\langle S\rangle\sigma\right)\eta_R^2  \nonumber \\
&+&\frac{1}{\sqrt 2}\left(\lambda_-\langle\phi\rangle h 
+\lambda_7\langle S\rangle\sigma\right)\eta_I^2  
+\frac{\lambda_6}{\sqrt 2}\left(\langle\phi\rangle h\sigma^2
+\langle S\rangle\sigma h^2\right) 
+  \frac{\lambda_5^\prime}{4\sqrt 2M_\ast}\sigma h^2(\eta_R^2-\eta_I^2),
\label{pot}
\end{eqnarray}
where we use the definition $\lambda_\pm=\lambda_3+\lambda_4\pm
\lambda_5$ and
\begin{equation}
M_{\eta_c}^2=m_\eta^2+\lambda_7\langle S\rangle^2 
+ \lambda_3\langle\phi\rangle^2, \qquad
M_{\eta_{R(I)}}^2=m_\eta^2+ \lambda_7\langle S\rangle^2 
+ \lambda_{+(-)}\langle\phi\rangle^2.
\label{emass}
\end{equation}
 The difference between these masses is estimated to be
\begin{equation}
\frac{M_{\eta_I}-M_{\eta_R}}{M_{\eta_R}}\simeq 
\frac{\lambda_5\langle\phi\rangle^2}{M_{\eta_R}^2}
\equiv\frac{\delta}{M_{\eta_R}}, \qquad
\frac{M_{\eta_c}-M_{\eta_R}}{M_{\eta_R}}\simeq 
\frac{(\lambda_4+\lambda_5)\langle\phi\rangle^2}{2M_{\eta_R}^2},
\label{mdif}
\end{equation}
which could be a good approximation as long as
$m_\eta^2+\lambda_7\langle S\rangle^2 \gg \langle\phi\rangle^2$ is
satisfied.  A large value of $m_\eta^2+\lambda_7\langle S\rangle^2$ is 
favored from the analysis of the $T$ parameter in precise 
measurements of the electroweak interaction \cite{idm,idm1}.
We assume such a situation in the present study.

Quartic scalar couplings in the potential $V$ are constrained by
several conditions. 
The stability of the assumed vacuum requires
\begin{equation}
\lambda_1,~ \lambda_2, ~\kappa >0; \quad \lambda_3,~\lambda_\pm >
 -2\sqrt{\lambda_1\lambda_2}; \quad \lambda_6>-2\sqrt{\lambda_1\kappa};
\quad \lambda_7>-2\sqrt{\lambda_2\kappa}.
\label{cstab}
\end{equation}
These can be easily read off from the expression of the scalar potential
$V$ given in
eq.~(\ref{pot}).\footnote{The last condition can be found by
using a different expression of $V$, which is modified so that 
$\sqrt\kappa$ has the opposite sign to eq.~(\ref{pot}).} 
Perturbativity of the model imposes that these quartic couplings 
should be smaller than $4\pi$.\footnote{More precisely, 
$|\lambda_{1,2}|$ and $|\kappa|$ should be smaller than $\frac{2\pi}{3}$.} 
Moreover, if we assume that $\eta_R$ is the lightest one among the fields 
with odd parity of the remnant $Z_2$, eq.~(\ref{emass}) shows
that the following conditions should be satisfied:
\begin{equation}
\lambda_4+\lambda_5<0, \qquad \lambda_5<0; \qquad 
M_{\eta_R} <{\rm min}(M_{\pm i}),
\label{cdm}
\end{equation}
where $M_{\pm i}$ are the mass eigenvalues for $N_i$ and $\tilde N_i$
which are discussed in detail later.
Using the value of $\lambda_1$ predicted by the Higgs mass 
observed at LHC experiments \cite{lhc} 
and the conditions given in eqs.~(\ref{cstab}) and (\ref{cdm}), 
we can roughly estimate the allowed range of $\lambda_{3,4}$ as
\begin{equation}
-2.5<\lambda_3<4\pi, \qquad -4\pi<\lambda_4 <0, 
\label{clambda}
\end{equation}
for sufficiently small values of $|\lambda_5|$.

The potential minimum in eq.~(\ref{pot}) is obtained as
\begin{equation}
\langle\phi\rangle^2=\frac{\lambda_6m_S^2-2\kappa
 m_\phi^2}{4\lambda_1\kappa-\lambda_6^2}, \qquad
\langle S\rangle^2=\frac{\lambda_6m_\phi^2-2\lambda_1
 m_S^2}{4\lambda_1\kappa-\lambda_6^2}.
\label{cvac}
\end{equation} 
Since the new gauge boson does not couple with the standard model fields,
both cases $\langle S\rangle^2 \gg\langle\phi\rangle^2$ and 
$\langle S\rangle^2 \ll \langle\phi\rangle^2$ could be
phenomenologically allowed. However, if we apply this model to the
leptogenesis, $\langle S\rangle^2 \gg\langle\phi\rangle^2$ should be
satisfied as discussed later.  
Such a vacuum can be realized for a sufficiently small $|\lambda_6|$ 
satisfying $4\lambda_1\kappa \gg \lambda_6^2$ and
negative values of $m_S^2$ and $m_\phi^2$ satisfying $|m_S^2|\gg |m_\phi^2|$.
In this case, both vacuum expectation values are approximately
expressed as
$\langle \phi\rangle^2\simeq -\frac{m_\phi^2}{2\lambda_1}$ and 
$\langle S\rangle^2\simeq -\frac{m_S^2}{2\kappa}$.
If the contribution of $\langle S\rangle$ to the $\eta$ mass
is of the same order as that of $\langle\phi\rangle$, 
$|\lambda_7|$ should be much smaller than $|\lambda_{3,4}|$ as found from
eq.~(\ref{emass}).

Since $h$ and $\sigma$ defined in eq.~(\ref{unitary}) 
have mass mixing as found from the first line in eq.~(\ref{pot}), 
mass eigenstates $\tilde h$ and $\tilde\sigma$ 
are a mixture of these. They are found to be written as
\begin{equation}
\tilde h\simeq h-\frac{\lambda_6\langle
 \phi\rangle}{2\kappa\langle S\rangle}\sigma, \qquad
\tilde\sigma=\sigma+\frac{\lambda_6\langle
 \phi\rangle}{2\kappa\langle S\rangle}h.
\end{equation}
However, since $\langle S\rangle^2\gg\langle \phi\rangle^2$ is 
assumed and $|\lambda_6|< \sqrt\kappa$ is expected, 
mass eigenstates could be almost equal 
to $h$ and $\sigma$.
In this case, the mass eigenvalues are approximately expressed as
\begin{equation}
m_{\tilde h}^2 = \left(4\lambda_1
-\frac{\lambda_6^2}{\kappa}\right)\langle\phi\rangle^2, \qquad
M_{\tilde\sigma}^2\simeq 4\kappa\langle S\rangle^2.
\label{hmass}
\end{equation}
These should have positive values for the stability of the
considered vacuum. It requires $4\lambda_1\kappa > \lambda_6^2$, 
which is consistent with the above discussion.

The value of $\lambda_1$ might be estimated by using
$m_{\tilde h}\simeq 125$~GeV.
If we apply it to the tree-level formula in eq.~(\ref{hmass}), we have 
\begin{equation}
\lambda_1-\frac{\lambda_6^2}{4\kappa}\sim 0.13. 
\label{lam1}
\end{equation}
This result suggests that $\lambda_1$ could have a somewhat 
larger value than the corresponding quartic coupling in the standard model.
However, this effect is expected to be small since the assumed vacuum 
requires $4\lambda_1\kappa \gg \lambda_6^2$.
On the other hand, the model has the additional 
scalar couplings $\lambda_3$ and $\lambda_4$, which are known to improve 
the potential stability \cite{stab}. 
Thus, the constraint from the potential stability against the radiative 
correction in the present model could be milder than that of 
the standard model. 

If we impose the requirement that $\tilde\sigma$ is heavier 
than the Higgs scalar, $\kappa$ satisfies 
$\kappa~{^>_\sim}~10^{-3}\left(\frac{2~{\rm TeV}}{\langle S\rangle}\right)^2$
and $\lambda_6$ could take a small value so as to be consistent with the 
condition $|\lambda_6|<2\sqrt{\lambda_1\kappa}$. 
If the above condition for $\kappa$ is not satisfied, 
$\tilde\sigma$ can be lighter than $\tilde h$ so as to realize 
$m_{\tilde h}>2 M_{\tilde\sigma}$.
In that case, the coupling $\lambda_6$ satisfies
$|\lambda_6|~{^<_\sim}~ 10^{-2}
\left(\frac{\lambda_1}{0.13}\right)^{\frac{1}{2}}
\left(\frac{2~{\rm TeV}}{\langle S\rangle}\right)$ and
the interaction in the last line of eq.~(\ref{pot}) 
induces the invisible decay $\tilde{h}\rightarrow 2\tilde\sigma$. 
The decay width can be estimated as
\begin{equation}
\Gamma(\tilde h\rightarrow 2\tilde\sigma)=
\frac{\lambda_6^2|\langle\phi\rangle|^2}
{16\pi m_{\tilde h}}\sqrt{1-4\frac{M_{\tilde\sigma}^2}{m_{\tilde h}^2}}.
\end{equation}
The branching ratio of this invisible decay should be less than $19\%$ 
of the Higgs total width $\sim 4$~MeV \cite{hwidth}. 
This constrains the value of $\lambda_6$ as $|\lambda_6|<0.0126$ \cite{wein},
which could be consistent with the vacuum condition discussed above.    
Here, we note that both $\kappa$ and $\lambda_6$ take small values for 
the light $\tilde\sigma$. In that case, $\tilde\sigma$ could have 
non-negligible cosmological effects.
We will come back to this point later.

\subsection{Degenerate right-handed neutrinos}
Next, we discuss the neutrino sector. 
If the thermal leptogenesis at TeV scales is supposed to be the origin 
of baryon number asymmetry in the Universe, the mass degeneracy 
among right-handed neutrinos is indispensable, at least 
in certain parameter regions \cite{ks}.
In the present model, spontaneous breaking of a new Abelian gauge 
symmetry due to a vacuum expectation value of $S$
could make the singlet fermions $N_i$ and $\tilde N_i$ behave as 
pseudo-Dirac fermions.
In fact, if $|y_i\langle S^\dagger\rangle|, 
|\tilde y_i\langle S\rangle|\ll M_i$ is satisfied, 
their masses are almost degenerate.\footnote{The same
scenario has been considered to explain the mass 
degeneracy among right-handed neutrinos first in \cite{res-f2}.
It is also discussed in \cite{pseudoD}.}

The mass matrix of the singlet fermions is expressed as
\begin{equation}
\frac{1}{2}(N_i, \tilde{N_i})\left(\begin{array}{cc}
|y_i|e^{i\gamma_i}\langle S^\dagger\rangle & M_i \\ M_i & 
|\tilde{y_i}|e^{i\tilde\gamma_i}
\langle S\rangle \\ 
\end{array}\right)
\left(\begin{array}{c} N_i \\ \tilde{N_i}\\\end{array}\right) +{\rm h.c.},
\end{equation}
where $M_i$ and $\langle S\rangle$ can be taken to be positive generally.
The mass eigenvalues $M_{\pm i}$  are derived as
\begin{eqnarray}
&& M_{+i}\simeq M_i\sin 2\theta_i+\Big(|y_i|\cos(\gamma_i-\xi_i)\cos^2\theta_i
+|\tilde y_i|\cos(\tilde\gamma_i+\xi_i)\sin^2\theta_i\Big)
\langle S\rangle, \nonumber \\
&& M_{-i}\simeq M_i\sin 2\theta_i -\Big(|y_i|\cos(\gamma_i-\xi_i)\sin^2\theta_i
+|\tilde y_i|\cos(\tilde\gamma_i+\xi_i)\cos^2\theta_i\Big)
\langle S\rangle,
\label{e-nmass}
\end{eqnarray}
and the corresponding mass eigenstates ${\cal N}_{\pm i}$ 
are found to be written as
\begin{eqnarray}
&&{\cal N}_{+i}=e^{-i\frac{\xi_i}{2}}
\left(N_i\cos\theta_i + \tilde{N_i}e^{-i\xi_i}\sin\theta_i\right),
\nonumber \\
&& {\cal N}_{-i}=ie^{-i\frac{\xi_i}{2}}
\left(-N_i\sin\theta_i + \tilde{N_i}e^{-i\xi_i}\cos\theta_i\right),
\end{eqnarray}  
respectively. Here, the phase $\xi_i$ is fixed by the parameters 
in the mass matrix as 
\begin{equation}
\tan\xi_i=\frac{|y_i|\sin\gamma_i-|\tilde y_i|\sin\tilde\gamma_i}
{|y_i|\cos\gamma_i+|\tilde y_i|\cos\tilde\gamma_i},
\label{beta}
\end{equation}
and the mixing angle $\theta_i$ is given by using this $\xi_i$ as
\begin{equation}
\tan2\theta_i=\frac{M_i}{\langle S\rangle}
\frac{2}{|y_i|\cos(\gamma_i-\xi_i) -|\tilde{y_i}|\cos(\tilde\gamma_i+\xi_i)}.
\label{theta}
\end{equation}
The difference of the mass eigenvalues given in eq.~(\ref{e-nmass}) 
is expressed by using these as
\begin{equation}
\Delta_i\equiv \frac{M_{+i}-M_{-i}}{M_{-i}}\simeq 
\frac{\langle S\rangle}{M_i}
\frac{|y_i|\cos(\gamma_i-\xi_i)+|\tilde y_i|\cos(\tilde\gamma_i +\xi_i)}
{\sin 2\theta_i}.
\label{msplit}
\end{equation}
From these formulas, we find that
$\theta_i$ could be approximated as $\frac{\pi}{4}$ and 
also the right-handed neutrino masses might be finely degenerate at a 
period where the sphaleron interaction is in thermal equilibrium,
simultaneously. The condition required for this is that 
both $|y_i|\langle S\rangle$ and $|\tilde y_i|\langle S\rangle$ are
much smaller than $M_i$ which is assumed to be of $O(1)$ TeV. 
This implies that the resonant leptogenesis could occur for 
a value of $\langle S\rangle$ which is larger than the weak scale
as long as both $|y_i|$ and $|\tilde y_i|$ are sufficiently small.  

The neutrino Yukawa couplings and other relevant
interactions of the right-handed neutrinos in eq.~(\ref{model}) 
can be written by using the mass eigenstates ${\cal N}_{\pm i}$ as
\begin{eqnarray}
&&\sum_{i=1,2}\Big[e^{-i\frac{\xi_i}{2}} 
\Big(h_{\alpha i}\cos\theta_i +\tilde h_{\alpha i}e^{-i\xi_i}\sin\theta_i\Big)
\bar{\cal N}_{+i}\eta^\dagger\ell_\alpha \nonumber \\ 
&&\hspace*{1cm}-ie^{-i\frac{\xi_i}{2}}\Big(h_{\alpha i}\sin\theta_i
-\tilde h_{\alpha i}e^{-i\xi_i}\cos\theta_i\Big)\bar{\cal N}_{-i}\eta^\dagger
\ell_\alpha  \nonumber \\
&&\hspace*{1cm}+\frac{1}{2\sqrt 2}\Big\{
\left(|y_i|e^{i(\gamma_i+\xi_i)}\cos^2\theta_i+|\tilde y_i|
e^{i(\tilde\gamma_i+ 3\xi_i)}\sin^2\theta_i\right)
\tilde\sigma{\cal N}_{+i}^2 \nonumber \\  
&&\hspace*{1cm}-\left(|y_i|e^{i(\gamma_i+\xi_i)}\sin^2\theta_i+|\tilde y_i|
e^{i(\tilde\gamma_i +3\xi_i)}\cos^2\theta_i\right)
\tilde\sigma{\cal N}_{-i}^2 \nonumber \\ 
&&\hspace*{1cm}+i\sin 2\theta_i\left(|y_i|e^{i(\gamma_i+\xi_i)}-|\tilde y_i|
e^{i(\tilde\gamma_i +3\xi_i)}\right)\tilde\sigma{\cal N}_{+i}{\cal N}_{-i}
\Big\} \nonumber \\
&&\hspace*{1cm}+ ig_X\sin 2\theta_i~X_\mu(\bar{\cal N}_{+i}
\gamma^\mu {\cal N}_{-i}) + {\rm h.c.}~ \Big].
\label{couplings}
\end{eqnarray}
If $h_{\alpha i}=\tilde h_{\alpha i}$ is satisfied,\footnote{Although this
assumption is not necessary for the present scenario, we adopt it to
make the analysis easier.}
the flavor structure of the model becomes very simple.
In that case, the neutrino Yukawa couplings can be rewritten as
\begin{eqnarray}
g_{\alpha i}^{(+)}&\equiv&e^{-i\frac{\xi_i}{2}} 
h_{\alpha i}\Big(\cos\theta_i +e^{-i\xi_i}\sin\theta_i\Big)
=h_{\alpha i}\left(1+\cos\xi_i\sin 2\theta_i\right)^{\frac{1}{2}}
e^{i(\delta_{+i}-\frac{\xi_i}{2})}, \nonumber \\
g_{\alpha i}^{(-)}&\equiv&-ie^{-i\frac{\xi_i}{2}}h_{\alpha i}
\Big(\sin\theta_i- e^{-i\xi_i}\cos\theta_i\Big)
=h_{\alpha i}\left(1-\cos\xi_i\sin 2\theta_i\right)^{\frac{1}{2}}
e^{i(\delta_{-i}-\frac{\xi_i}{2})},
\label{nyukawa}
\end{eqnarray}
where we suppose $h_{\alpha i}$ to be real, for simplicity.
The phases $\delta_{\pm i}$ are defined as
\begin{equation}
\tan\delta_{+ i}=\frac{-\sin\xi_i\tan\theta_i}
{1+\cos\xi_i\tan\theta_i}, \qquad
\cot\delta_{-i}=\frac{\sin\xi_i}{\cos\xi_i-\tan\theta_i}.
\end{equation}
We use these simplified neutrino Yukawa couplings 
in the following discussion. 

The neutrino mass is induced through one-loop diagrams which 
have ${\cal N}_{+i}$ or ${\cal N}_{-i}$ in an internal fermion line
as in the original model.
The mass formula is given by
\begin{equation}
{\cal M}_{\alpha\beta}
=\sum_{i}\sum_{s=\pm}\left|g_{\alpha i}^{(s)}g_{\beta i}^{(s)}\lambda_5\right|
e^{i(2\delta_{si}-\xi_i)}\Lambda(M_{s i}),
\label{nmass}
\end{equation}
where $\Lambda(M_{\pm i})$ is defined as
\begin{equation}
\Lambda(M_{\pm i})=\frac{\langle\phi\rangle^2}{8\pi^2}
\frac{M_{\pm i}}{M_\eta^2-M_{\pm i}^2}
\left(1+\frac{M_{\pm i}^2}{M_\eta^2-M_{\pm i}^2}
\ln\frac{M_{\pm i}^2}{M_\eta^2}\right).
\label{nmass1}
\end{equation}
$M_\eta$ is an averaged value of the mass eigenvalues of $\eta_R$ and $\eta_I$.
If the model has two sets of $(N_i, \tilde N_i)$ at least,
neutrino mass eigenvalues suitable for the explanation of 
the neutrino oscillation data 
could be derived.\footnote{We can consider another minimal 
model which has one set of $(N_1, \tilde N_1)$ and an additional
right-handed neutrino which has no charge of U(1)$_X$. A result similar 
to the present one could be expected for neutrino masses and
leptogenesis also in such a model.}
We consider a model with two sets of $(N_i, \tilde N_i)$ in the following.

Since the scale $\Lambda(M_{\pm i})$ is estimated as
$\Lambda(M_{\pm i})=O(10^9)$~eV for $\eta$ and ${\cal N}_{\pm i}$ 
whose masses are in the TeV range, eq.(\ref{nmass}) suggests that 
the atmospheric neutrino data require the relevant neutrino Yukawa couplings 
to satisfy
\begin{equation}
\sum_i \left|g_{\alpha i}^{(\pm)}g_{\beta i}^{(\pm)}\lambda_5\right|
=O(10^{-11}).
\label{c-nmass}
\end{equation}
On the other hand, if ${\cal N}_{-1}$ is identified with the lightest 
right-handed neutrino, its decay should occur in out-of-thermal 
equilibrium for successful leptogenesis. 
This condition could impose strong constraints on various interactions
of ${\cal N}_{-1}$. They can be roughly estimated by imposing both reaction 
rates of the decay of ${\cal N}_{-1}$ and its scattering with 
other particles to be smaller than the Hubble parameter.
The most important process is the ${\cal N}_{-1}$ decay. 
If the neutrino Yukawa couplings of ${\cal N}_{-1}$ satisfy 
\begin{equation}
\left(\sum_\alpha \left|g_{\alpha 1}^{(-)}\right|^2\right)^{\frac{1}{2}} 
\le 10^{-8},
\label{lept1}
\end{equation}
it does not reach equilibrium at the temperature $T~{^>_\sim}~100$~GeV.    

The condition (\ref{lept1}) shows that ${\cal N}_{-1}$ causes a 
negligible contribution to the neutrino mass generation, which is found 
from eqs.~(\ref{nmass}) and (\ref{c-nmass}). On the other hand,  
if ${\cal N}_{+1}$ is supposed to cause a main contribution to
the neutrino mass generation, the condition (\ref{c-nmass}) shows that
its Yukawa couplings should satisfy
\begin{equation}
\left|g_{\alpha 1}^{(+)}\right|^2= 
O\left(\frac{10^{-11}}{|\lambda_5|}\right)
\qquad (\alpha=e,\mu,\tau).
\label{lept2}
\end{equation}
Equation (\ref{nyukawa}) suggests that
the original neutrino Yukawa couplings $|h_{\alpha 1}|$ do not need to be 
extremely small for the simultaneous realization of the conditions
 (\ref{lept1}) and (\ref{lept2}) 
as long as $\cos\xi_1\sin 2\theta_1\simeq 1$ is satisfied
to a good accuracy and also $|\lambda_5|$ takes a small value of
$O(10^{-4})$. 
Other nonzero neutrino mass eigenvalues could be determined 
through the second pair $(N_2,\tilde N_2)$.
Since the relevant Yukawa couplings $h_{\alpha 2}$ are not constrained 
by the leptogenesis, we can derive neutrino masses 
and mixing favorable for the explanation 
of the neutrino oscillation data through eq.~(\ref{nmass}) 
independently.
If only one of ${\cal N}_{\pm 2}$ contributes to the neutrino mass
generation as in the $(N_1, \tilde N_1)$ sector, one of three
neutrino mass eigenvalues is expected to be negligibly small as in the
model studied in \cite{ks}.  

\input epsf
\begin{figure}[t]
\begin{center}
\epsfxsize=7.5cm
\leavevmode
\epsfbox{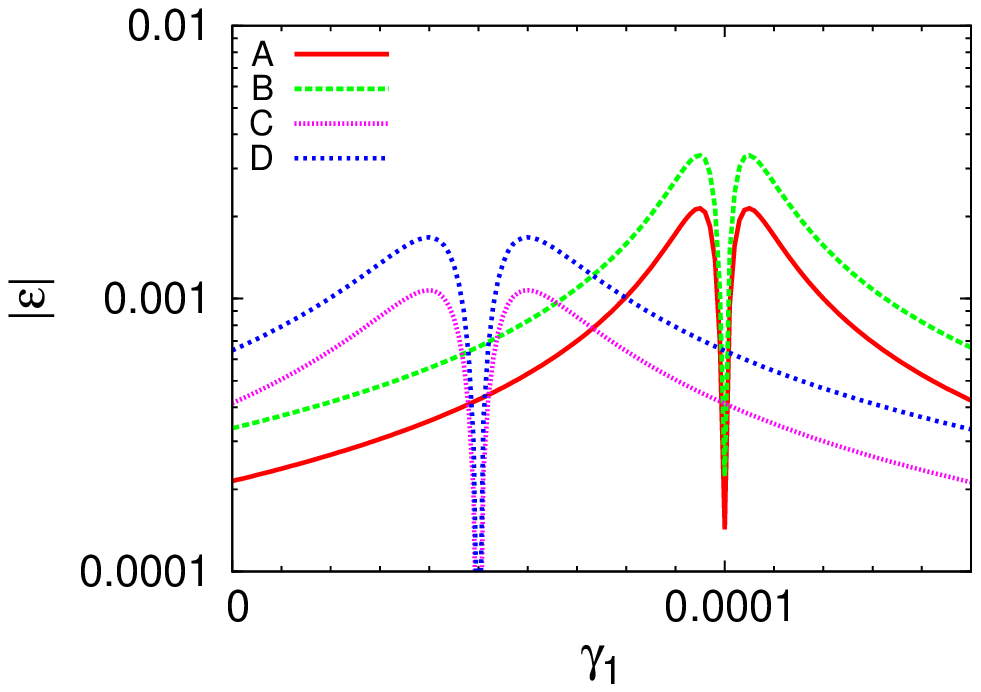}
\end{center}
\vspace*{-3mm}

{\footnotesize {\bf Fig.~1}~~$CP$ asymmetry as a function of
 $\gamma_1$ for typical values of $(|y_1|, |h_{\alpha 1}|)$.
In each case, these parameters are fixed as 
A$(10^{-5}, 4\times 10^{-4})$,
B$(10^{-5}, 5\times 10^{-4})$, 
C$(2\times 10^{-5}, 4\times 10^{-4})$, and 
D$(2\times 10^{-5}, 5\times 10^{-4})$.
Other relevant parameters are taken to be $\tilde\gamma_1=0.1$, 
$\tilde y_1=10^{-8}$, $M_1=\langle S\rangle=2$~TeV and $M_\eta=1$~TeV.}  
\end{figure}

\subsection{Resonant leptogenesis}
In this framework, we consider resonant leptogenesis \cite{res,res-f1,res-f2}.  
The dominant contribution to the $CP$ asymmetry $\varepsilon$ 
in the ${\cal N}_{-1}$ decay comes from the resonance appearing in the 
one-loop self-energy diagram. In that case, $\varepsilon$ is known to be
expressed as \cite{res-f1,res-f2}
\begin{eqnarray}
\varepsilon&=&\frac{{\rm Im}
\left(\sum_\alpha g_{\alpha 1}^{(+)\ast} g_{\alpha 1}^{(-)}\right)^2}
{\left(\sum_\alpha g_{\alpha 1}^{(-)\ast} g_{\alpha 1}^{(-)}\right)
\left(\sum_\alpha g_{\alpha 1}^{(+)\ast}g_{\alpha 1}^{(+)}\right)}~
\frac{2\Delta_1{\tilde\Gamma_{{\cal N}_{+1}}}}
{4\Delta_1^2+\tilde\Gamma_{{\cal N}_{+1}}^2} \nonumber \\
&=&\frac{\cos2\theta_1\sin
 2\xi_1}{1-\sin^22\theta_1\cos^2\xi_1}
\frac{2\Delta_1{\tilde\Gamma_{{\cal N}_{+1}}}}
{4\Delta_1^2+\tilde\Gamma_{{\cal N}_{+1}}^2},
\end{eqnarray}
where we use the expression of the neutrino Yukawa couplings 
$|g_{\alpha 1}^{\pm}|$ given in eq.~(\ref{nyukawa}).
The mass degeneracy $\Delta_1$ is defined in eq.~(\ref{msplit}) 
and $\tilde\Gamma_{{\cal N}_{+1}}=
\frac{\sum_\alpha\left|g_{\alpha 1}^{(+)}\right|^2} 
{8\pi}\left(1-\frac{M_\eta^2}{M_{+1}^2}\right)^2$.
If we assume $\langle S\rangle=M_1$ for simplicity, the right-handed
neutrino sector $(N_1,\tilde N_1)$ has five free parameters.  
Using these, we study the relation between the $CP$ asymmetry 
and the structure of right-handed neutrino sector. 

In Fig.~1, we plot the $CP$ asymmetry $\varepsilon$ as a function
of $\gamma_1$ for four typical sets of $(|y_1|, |h_{\alpha 1}|)$.
Other parameters are fixed at the values given in the caption of Fig.~1. 
We find that $\varepsilon$ changes the sign from minus to plus at 
$\gamma_1\sim 10^{-4}$ and $5\times 10^{-5}$ for the cases A, B 
and C, D, respectively. Its absolute value is enhanced largely 
around these values of $\gamma_1$.
If we note that $\left|g_{\alpha 1}^{(-)}\right|\le O(10^{-8})$ is 
required for the out-of-equilibrium decay of ${\cal N}_{-1}$, 
we find that $|\xi_1|$ should take 
a very small value such as $O(10^{-4})$ for $|h_{\alpha 1}|=O(10^{-4})$.
As found from eq.~(\ref{beta}), such a small $|\xi_1|$ 
could be easily realized for hierarchical $|y_1|$ and $|\tilde y_1|$
by fixing the values of $\gamma_1$ and $\tilde\gamma_1$ appropriately.
In these examples, such hierarchical values  
are assumed for $|y_1|$ and $|\tilde y_1|$. 
We also note that the same parameter set could induce the degenerate 
right-handed neutrino masses as found from eq.~(\ref{msplit}).
This feature makes it for the model possible to satisfy the minimum conditions
for the success of resonant leptogenesis. 
Although we have to introduce a tiny coupling $|\tilde y_1|$ in this scenario,
the important quantities for the leptogenesis are closely related each other.
The model can bring about their favorable values simultaneously 
based on the common parameters.
In fact, for the parameters used in Fig.~1, the desirable values of 
the relevant quantities to the leptogenesis can be obtained. 
We present their values derived 
from these parameters in Table 1.
These results show that $\left|g_{\alpha 1}^{(-)}\right|$ takes small values which 
satisfy the condition (\ref{lept2}) at the points where the $CP$ 
asymmetry $|\varepsilon|$ has large values.
The mass degeneracy $\Delta_1=O(10^{-5})$ between the right-handed 
neutrinos ${\cal N}_{\pm 1}$ is also realized at this region. 
This level of degeneracy has been shown to be sufficient 
for the leptogenesis in the radiative neutrino mass model in the previous
study \cite{ks}. 
Although the smallness of $|\tilde y_1|$ should be explained by considering 
some complete model in the high energy region, it is beyond the
scope of the present study and we do not go further in this direction here.
 
\begin{figure}[t]
\begin{center}
\begin{tabular}{ccccc}\hline
 & $|g_{\alpha 1}^{(-)}|$ & 
$|g_{\alpha 1}^{(+)}|$ & $\Delta_1$ &$\varepsilon$ \\ \hline\hline
A  & $3.12\cdot 10^{-9}$ & $5.66\cdot 10^{-4}$ 
& $1.00\cdot 10^{-5}$ & $-1.73 \cdot 10^{-3}$  \\
B & $6.71\cdot 10^{-9}$ & $7.07\cdot 10^{-4}$ 
& $1.00\cdot 10^{-5}$ & $-2.71\cdot 10^{-3}$ \\
C & $1.17\cdot 10^{-8}$ & $5.66\cdot 10^{-4}$ 
& $2.00\cdot 10^{-5}$ & $-5.04\cdot 10^{-4}$ \\
D &$1.46\cdot 10^{-8}$ & $7.07\cdot 10^{-4}$ 
& $2.00\cdot 10^{-5}$ & $-7.88\cdot 10^{-4}$  \\ \hline
\end{tabular}
\end{center}
\vspace*{-1mm}

{\footnotesize {\bf Table~1}~ Derived values of the quantities
relevant to the leptogenesis for each case given in Fig.~1. 
These are estimated at $\gamma_1\sim 9 \times 10^{-5}$ 
and $4\times 10^{-5}$ for the cases A, B and C, D, respectively. }  
\end{figure}

The baryon number asymmetry generated through the decay of ${\cal N}_{-1}$ 
can be fixed by estimating the generated lepton number asymmetry
through solving the Boltzmann equations numerically for both
the ${\cal N}_{-1}$ number density $n_{{\cal N}_{-1}}$ and the lepton
number asymmetry $n_L(\equiv n_\ell-n_{\bar\ell})$. 
We introduce these number densities in the co-moving volume 
as $Y_{{\cal N}_{-1}}=\frac{n_{{\cal N}_{-1}}}{s}$ 
and $Y_L=\frac{n_L}{s}$ by using the entropy density $s$. 
The Boltzmann equations for these are written as 
\begin{eqnarray}
&&\frac{dY_{{\cal N}_{-1}}}{dz}=-\frac{z}{sH(M_{-1})}
\left(\frac{Y_{{\cal N}_{-1}}}{Y_{{\cal N}_{-1}}^{\rm eq}}-1\right)\left\{
\gamma^D_{{\cal N}_{-1}}+ \gamma_{{{\cal N}_{-1}}
{\tilde\sigma}}^{S}+ \gamma_{{{\cal N}_{-1}}{X}}^{S}\right\}, 
\nonumber \\
&&\frac{dY_L}{dz}=\frac{z}{sH(M_{-1})}\left\{
\varepsilon\left(\frac{Y_{{\cal N}_{-1}}}{Y_{{\cal N}_{-1}}^{\rm eq}}-1\right)
\gamma^D_{{\cal N}_{-1}}-\frac{2Y_L}{Y_\ell^{\rm eq}}
\left(\frac{\gamma^D_{{\cal N}_{+1}}}{4}
+\gamma_{{\cal N}_{+1}}^{(2)} +\gamma_{{\cal N}_{+1}}^{(13)}\right)\right\}, 
\label{bqn}
\end{eqnarray}
where $z=\frac{M_{-1}}{T}$ and $H(M_{-1})=1.66g_\ast^{1/2}\frac{M_{-1}^2}
{m_{\rm pl}}$. The equilibrium values for these are expressed as 
$Y_{{\cal N}_{-1}}^{\rm eq}(z)=\frac{45}{2\pi^4g_\ast}z^2K_2(z)$
and $Y_\ell^{\rm eq}\simeq\frac{81}{\pi^4g_\ast}$, where $K_2(z)$ is the
modified Bessel function of the second kind.
Since the Yukawa couplings of ${\cal N}_{+1}$ are large enough,
it is expected to be in thermal equilibrium throughout the relevant
period.
In these equations, we take into account the important reactions
which could keep ${\cal N}_{-1}$ in the equilibrium and wash out the
generated lepton number asymmetry. 
The former ones include the 2-2 scatterings of ${\cal N}_{-1}$ with 
$\tilde\sigma$ and $X_\mu$, whose reaction densities are represented 
by $\gamma^S_{{\cal N}_{-1}\tilde\sigma}$ and 
$\gamma^S_{{\cal N}_{-1}X}$ in eq.~(\ref{bqn}). 
These could be effective if $\tilde\sigma$ and $X_\mu$ are light enough.
Other reaction densities in eq.~(\ref{bqn}) can be found in the appendix 
of \cite{ks}.

\begin{figure}[t]
\begin{center}
\epsfxsize=4.8cm
\leavevmode
\epsfbox{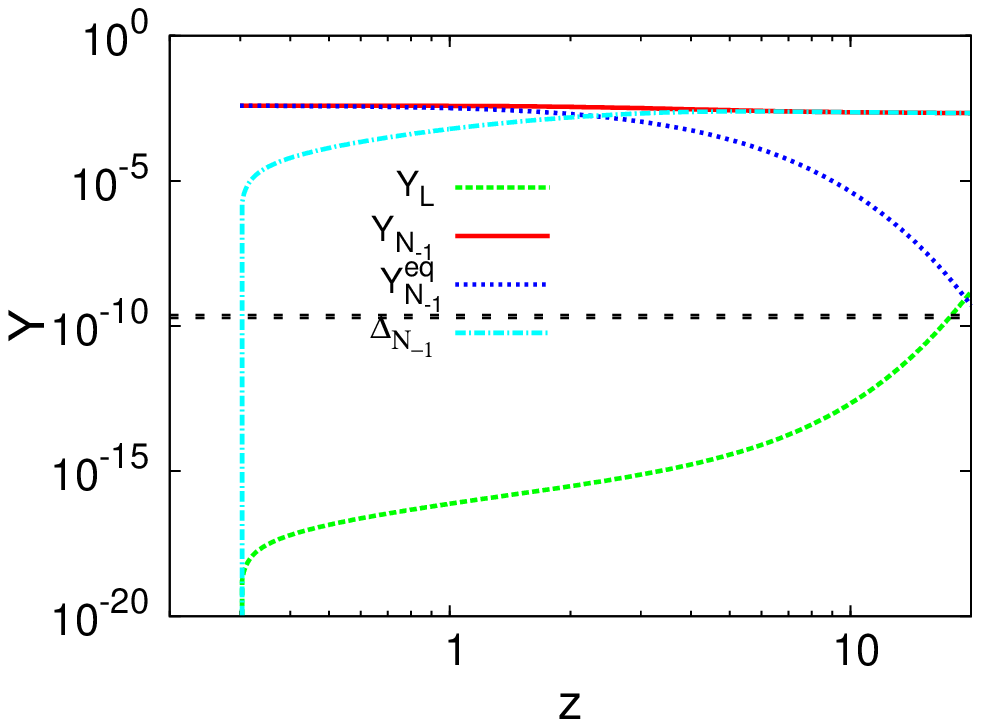}
\hspace*{3mm}
\epsfxsize=4.8cm
\leavevmode
\epsfbox{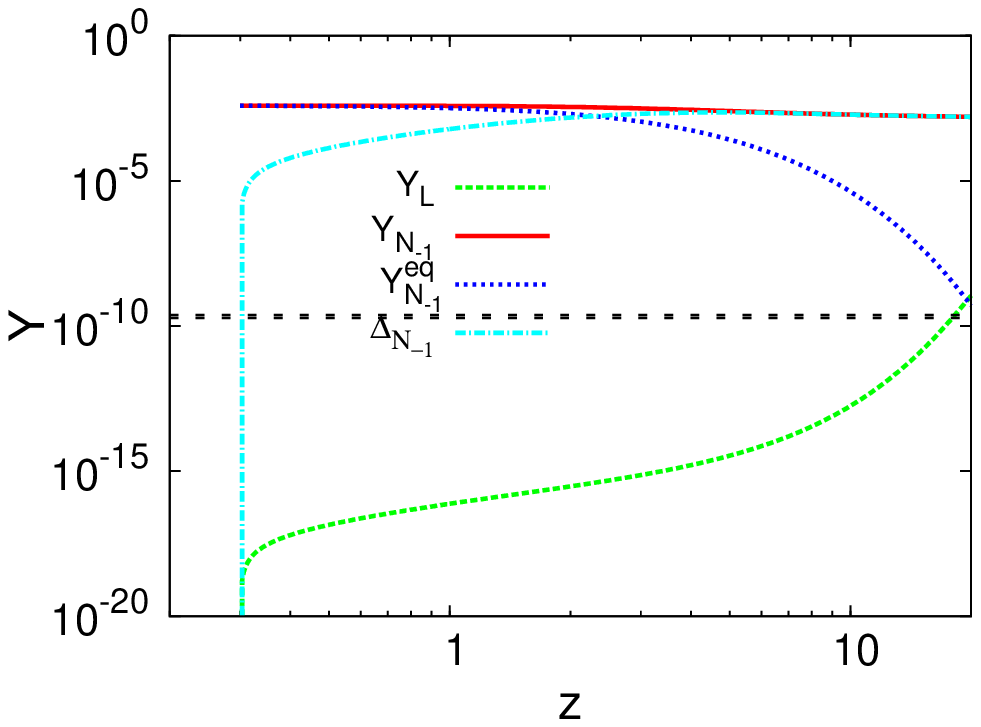}
\hspace*{3mm}
\epsfxsize=4.8cm
\leavevmode
\epsfbox{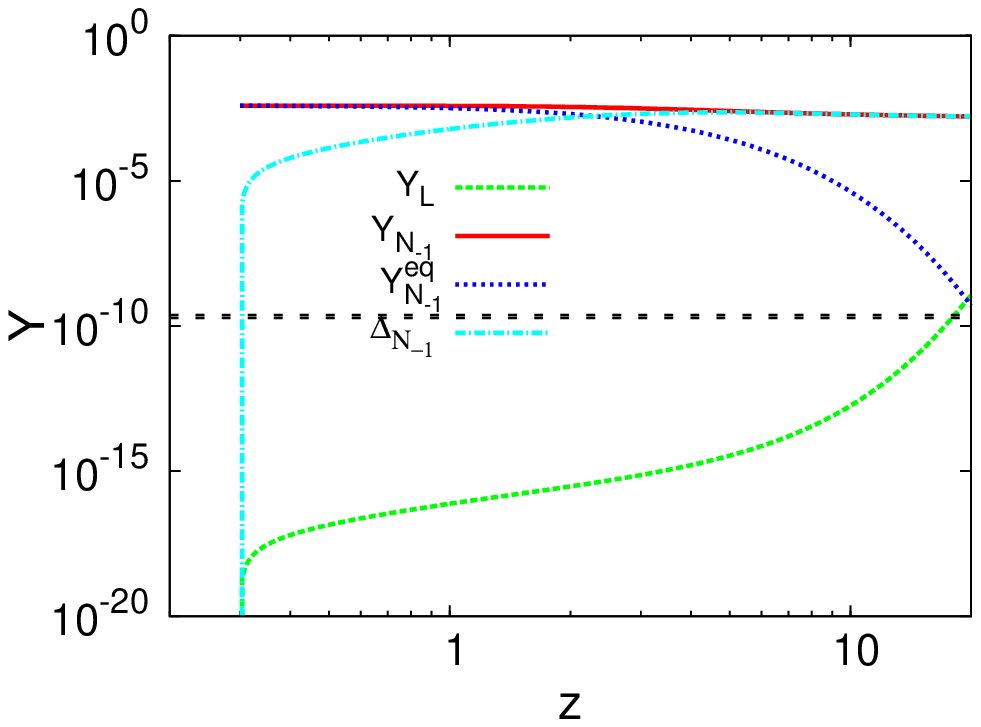}
\\
\epsfxsize=4.8cm
\leavevmode
\epsfbox{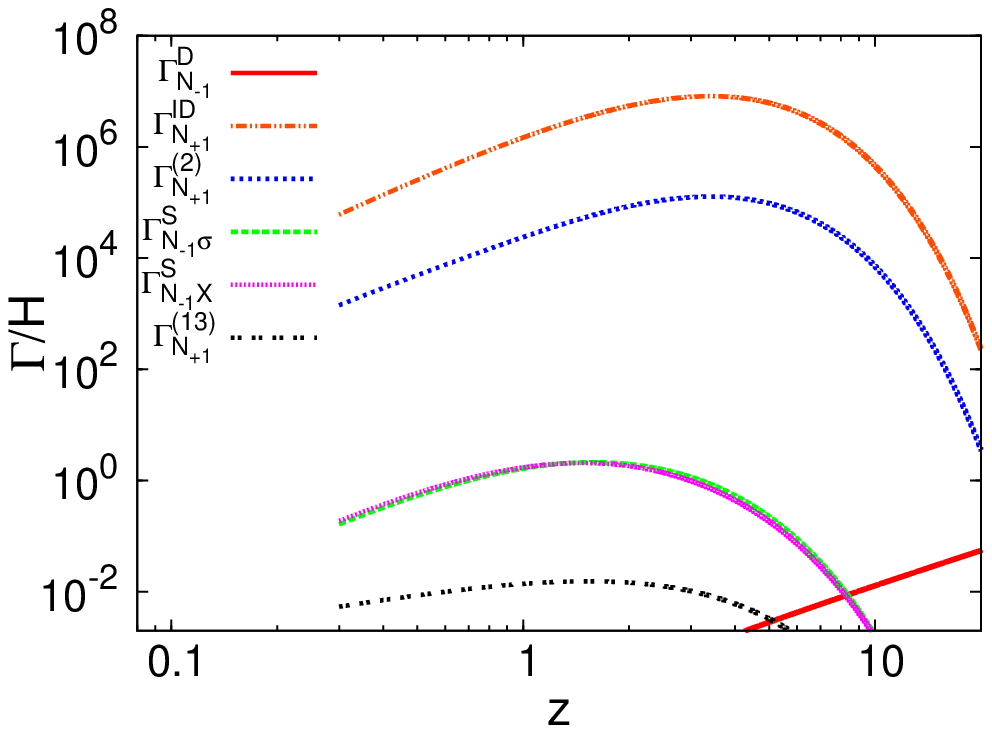}
\hspace*{3mm}
\epsfxsize=4.8cm
\leavevmode
\epsfbox{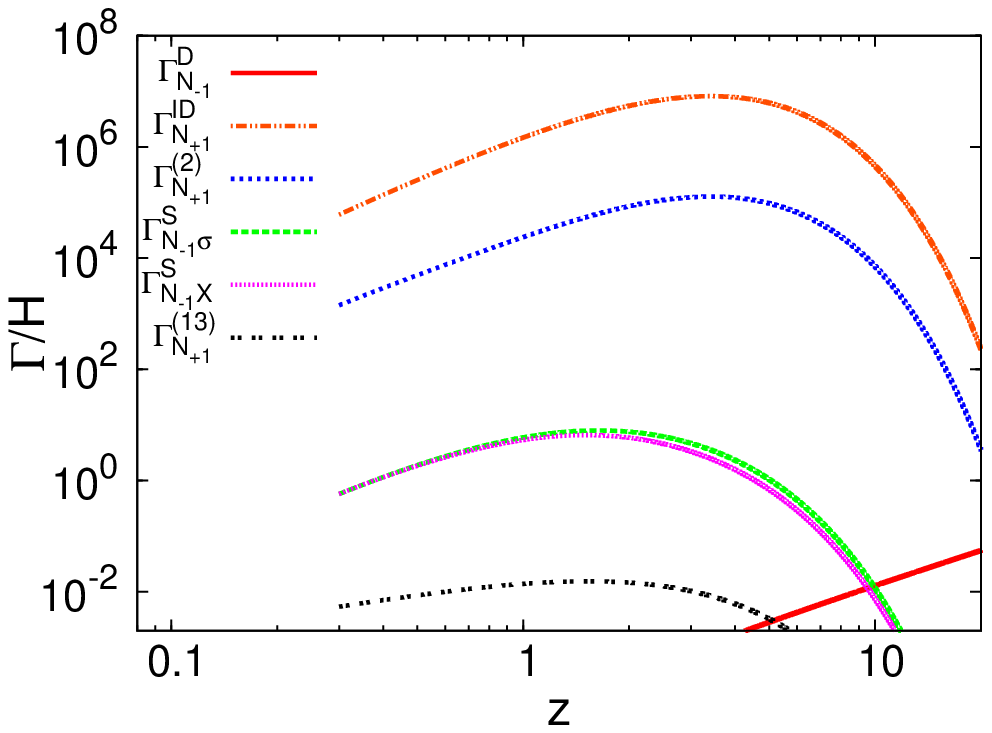}
\hspace*{3mm}
\epsfxsize=4.8cm
\leavevmode
\epsfbox{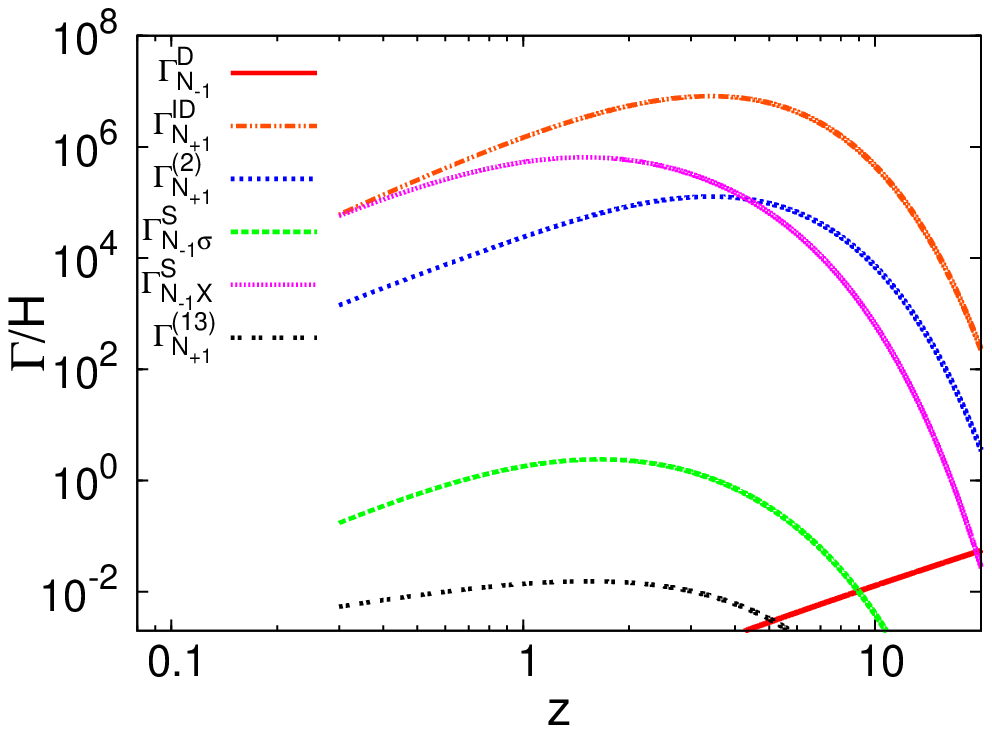}
\\
\end{center}
\vspace*{-3mm}

{\footnotesize {\bf Fig.~2}~~In the upper panels, solutions of 
the Boltzmann equations are plotted as a function of $z$ for the case A 
shown in Table~1. 
 In the lower panels, relevant reaction rates $\Gamma/H$ are plotted 
as a function of $z$ for the same parameters used in 
the corresponding upper panels. Reaction rates of the ${\cal N}_{-1}$ decay, 
the ${\cal N}_{+1}$ inverse decay and the lepton number violating 
${\cal N}_{+1}$ scatterings are represented by $\Gamma_{N_{-1}}^D$, 
$\Gamma_{N_{+1}}^{ID}$ and $\Gamma_{N_{+1}}^{(2)}$, $\Gamma_{N_{+1}}^{(13)}$, 
respectively. Masses of $\tilde\sigma$ and $X_\mu$ are set as 
$(M_{\tilde\sigma},m_X)=(200,300)$, $(60,100)$ and $(200,10^{-3})$ in a GeV unit 
from left to right, respectively.}  
\end{figure}

In Fig.~2, the solutions of these equations and the reaction
rates $\Gamma$ of the relevant processes are plotted as functions of $z$
for the case A in Table 1. In these panels, the masses of $\tilde\sigma$ and
$X_\mu$ are fixed to be $(M_{\tilde\sigma}, m_X)=(200,300)$, $(60, 100)$, 
and $(200, 10^{-3})$ in GeV units, respectively. 
As the initial condition for $Y_{{\cal N}_{-1}}$ in the Boltzmann 
equations we use its equilibrium value, since both $N_1$ and $\tilde N_1$
are expected to be in thermal equilibrium. 
Since we adopt this initial condition,
its deviation $\Delta_{{\cal N}_{-1}}$ from the equilibrium value does
not change sign as found in the upper panels of this figure. 
After $\langle S\rangle$ becomes nonzero, the mass eigenstate 
${\cal N}_{-1}$ leaves the equilibrium because of its small Yukawa coupling 
$g_{\alpha 1}^{(-)}$. Thus, it could be crucial in 
the estimation of the lepton number asymmetry at what value of $z$ we introduce
the effect of nonzero $\langle S\rangle$ in the equations. 
As a simple approximation, we introduce its effect as a step function at
$z_0$. In order to check the validity of this analysis, we change the
value of $z_0$ in the range $0.3<z_0<1$ to examine the $z_0$
dependence of the final results. 
Since their difference stays at most in a few 10\% range without showing 
a serious $z_0$ dependence, the present treatment can be considered 
to give reliable results. 

In the lower panels, which plot the behavior of the reaction rates, 
we find that the inverse decay of ${\cal N}_{+1}$ 
plays a dominant role for the wash-out of the generated lepton number 
asymmetry among various processes. Although the ${\cal N}_{+1}$ mass 
is almost degenerate with the mass of ${\cal N}_{-1}$,  its Yukawa coupling 
$g_{\alpha 1}^{(+)}$ is not so small as to decouple at an earlier period. 
This is an expected feature in the resonant leptogenesis generally.
The rapid increase of the lepton number asymmetry shown in the $z>10$ 
region can be understood from the large decrease of $\Gamma_{N_{+1}}^{ID}$ there.
The scatterings of ${\cal N}_{-1}$ with $\tilde\sigma$ and $X_\mu$ 
cannot be effective in keeping ${\cal N}_{-1}$ in thermal equilibrium
even if $\tilde\sigma$ and $X_\mu$ are light enough.  
Since $\langle S\rangle$ is supposed to be rather large, the assumed masses 
for $\tilde\sigma$ and $X_\mu$ are obtained only for the small couplings 
$\kappa$ and $g_X$. This is considered to be the cause of these results.

\begin{figure}[t]
\begin{center}
\begin{tabular}{c|ccccc}\hline
 $(M_{\tilde\sigma},~m_X)$ & A & B & C & D \\ \hline\hline
 $(200,~300)$ & $5.2\cdot 10^{-10}$ & $2.3\cdot 10^{-9}$ &
 $4.2\cdot 10^{-10}$ & $5.6 \cdot 10^{-10}$  \\ 
 $(60, ~100)$ & $3.9\cdot 10^{-10}$ & $1.7\cdot 10^{-9}$ &
 $1.5\cdot 10^{-10}$ & $1.9\cdot 10^{-10}$  \\
 $(200,~10^{-3})$ &  $4.0 \cdot 10^{-10}$ & $1.8 \cdot 10^{-9}$ &
 $1.6\cdot 10^{-10}$ & $2.2\cdot 10^{-10}$  \\
 $(600,~600)$ & $7.0 \cdot 10^{-10}$ & $ 3.1\cdot 10^{-9}$ & 
$1.1\cdot 10^{-9}$ &  $1.4\cdot 10^{-9}$  \\
  \hline
\end{tabular}
\end{center}
\vspace*{-1mm}

{\footnotesize {\bf Table~2}~ Baryon number asymmetry $Y_B$ predicted 
for the parameter sets given in Table 1. $M_{\tilde\sigma}$ and $m_X$ are
given in GeV units.}  
\end{figure}

The baryon number asymmetry $Y_B(\equiv\frac{n_B}{s})$ 
is expressed by using the solution $Y_L$ of the Boltzmann
equations as
\begin{equation}
Y_B=-\frac{8}{23}Y_L(z_{\rm EW}),
\end{equation}
where $z_{\rm EW}$ is related to the sphaleron decoupling temperature
$T_{\rm EW}$ by $z_{\rm EW}=\frac{M_{-1}}{T_{\rm EW}}$.
The baryon number asymmetry predicted for the parameters given in Table 1 
is listed in Table 2 for several values of $(M_{\tilde\sigma},m_X)$.      
These results show that the model could generate the sufficient baryon
number asymmetry compared with 
$8.1\times 10^{-11}<Y_B< 9.2\times 10^{-11}~(95\% CL)$ required from 
the observation \cite{pdg} as long as 
the relevant parameters take suitable values.\footnote{For more precise 
estimation, one could refer to the study in \cite{res2}, which 
includes the analysis not only for the phenomenon of mixing of 
heavy neutrinos, but also for oscillations among the heavy neutrinos.}
We note that the light $\tilde\sigma$ which can contribute to the 
invisible decay of the Higgs particle $\tilde h$ is also
allowed from the view point of 
the generation of baryon number asymmetry. 

The condition (\ref{c-nmass}) imposed by the neutrino oscillation data
requires $|\lambda_5|=O(10^{-4})$ for the above numerical results. 
As we will see in the next section, it is consistent with the constraint 
derived from the dark matter direct search.
The values of $\lambda_5$ and $\tilde h_{\alpha 1}$ used in the above 
study are found to be realized through eq.~(\ref{l5c}) for the cut-off 
scale such as $M_\ast=O(10^4)$~TeV, since we assume 
$\langle S\rangle =M_1$ here. 
Even if we do not assume this relation and 
$\langle S\rangle$ is supposed to have a larger value, 
a similar result is expected to be obtained for a larger value 
of $M_\ast$ and smaller values of $|y_i|$ and $|\tilde y_i|$.
  
\section{Physics in dark sector}
\subsection{Relic abundance and detection of dark matter}
It is well known that there are three possible mass ranges for an inert
doublet dark matter to realize the required relic abundance \cite{idm,idm1}. 
We are considering the high mass possibility 
here.\footnote{We note that a much more severe mass
degeneracy between the right-handed neutrinos is required
in the low mass possibility if the resonant
leptogenesis is applied to the model. This is because the wash-out of 
the generated lepton asymmetry is kept in the thermal equilibrium 
until a much later period in this case.}  
The $\eta_R$ relic abundance can be estimated along the same lines as the
original model \cite{ks,idm1}. However, we have to take into account that 
 the thermally averaged (co)annihilation cross section 
$\langle\sigma_{\rm eff}v\rangle$ has additional contributions 
from the processes which have $X_\mu$ or $\tilde\sigma$ in the
final states or intermediate states in the present model.
Moreover, for the inert doublet dark matter $\eta_R$, 
the direct search imposes severe constraints on the scalar couplings 
$\lambda_i$.

First, we consider the constraint induced through inelastic
scattering of $\eta_R$ with a nucleus.
Since the masses of $\eta_R$ and $\eta_I$ 
are almost degenerate for the small values of $|\lambda_5|$ as found from
eq.~(\ref{mdif}), this inelastic scattering of $\eta_R$ 
mediated by the $Z^0$ exchange brings about substantial 
effects to the direct search experiments \cite{inel,l5}. 
The interaction of $\eta_{R}$ relevant to this process is given 
by
\begin{equation}
{\cal L}=\frac{g}{2\cos\theta_W}Z^\mu\left(\eta_R\partial_\mu\eta_I 
- \eta_I\partial_\mu\eta_R\right).
\label{v-int}
\end{equation}
Inelastic $\eta_R$-nucleus scattering can occur for $\eta_R$ whose velocity 
is larger than the minimum value \cite{inelvel} given by 
\begin{equation}
v_{\rm min}=\frac{1}{\sqrt{2m_NE_R}}\left(\frac{m_NE_R}{\mu_N}
+\delta\right),
\end{equation}
where $\delta$ is the mass difference between $\eta_R$ and $\eta_I$ 
defined in eq.(\ref{mdif}). $E_R$ is the nucleus recoil
energy. The mass of the target nucleus and the
reduced mass of the nucleus-$\eta_R$ system are 
represented by $m_N$ and $\mu_N$.
The mass difference $\delta$ is constrained by the fact that 
no dark matter signal has been found in the direct search 
yet \cite{direct1,lux,direct2}.
This condition might be estimated as $\delta~{^>_\sim}150$~keV \cite{l5}. 
Since $\delta$ is related to $\lambda_5$ through eq.~(\ref{mdif}), 
the condition on $\delta$ constrains the value of $|\lambda_5|$ 
to satisfy \cite{ks}
\begin{equation}
|\lambda_5|\simeq \frac{M_{\eta_R}\delta}{\langle\phi\rangle^2} 
~{^>_\sim}~5.0\times 10^{-6}
\left(\frac{M_{\eta_R}}{1~{\rm TeV}}\right)
\left(\frac{\delta}{150~{\rm keV}}\right).
\label{direct}
\end{equation}
Since $\tilde\lambda_5=O(1)$ is expected, 
eq.~(\ref{l5c}) suggests that 
$\langle S\rangle~{^>_\sim}~5\times 10^{-6}M_\ast$ should be
satisfied.

The present results from a dark matter direct search also impose a constraint
on the values of the scalar couplings $\lambda_{3,4}$ and $\lambda_6$.
The elastic scattering $\eta_R$-nucleus is induced through the exchange of 
$\tilde h$ and $\tilde\sigma$.
The corresponding cross section for $\eta_R$-nucleon scattering at zero
momentum transfer can be calculated to be
\begin{equation}
\sigma_n^0=\frac{f^{(n)2}m_n^4\lambda_+^2}{8\pi M_{\eta_R}^2m_{\tilde h}^4}
\left(1 +\frac{\lambda_6^2}{4\kappa\lambda_1}\right)^2,
\label{direct0}
\end{equation}
where $m_n$ is a nucleon mass and $f^{(n)}\simeq 0.3$.
The second term in the parentheses comes from the $\tilde\sigma$ exchange.
If we apply the present direct search constraint  
$\sigma_n^0<1\times 10^{-44}~{\rm cm}^2$ for $M_{\eta_R}=O(1)$~TeV 
\cite{lux},
we find that the scalar couplings $\lambda_{3,4}$ should satisfy
\begin{equation}
\lambda_+\left(1+\frac{\lambda_6^2}{4\kappa\lambda_1}\right)
<1.5\left(\frac{M_{\eta_R}}{1~{\rm TeV}}\right),
\label{direct01}
\end{equation} 
where $\lambda_+\simeq\lambda_3+\lambda_4$.
Since the potential stability requires $\lambda_6^2<4\kappa\lambda_1$ 
as seen before, the $\tilde\sigma$ exchange contribution to the
$\eta_R$-nucleon scattering can be generally neglected
except for the case where $\lambda_6^2$ takes the same value
as regards the order, $4\kappa\lambda_1$. 

We now proceed to the estimation of the $\eta_R$ relic abundance taking
account of the conditions discussed above. We use the
notation $(\eta_1,\eta_2,\eta_3,\eta_4)=(\eta_R,\eta_I,\eta_+,\eta_-)$
for convenience here.
The dominant parts of the effective (co)annihilation cross section 
including the new contributions are calculated to be
\begin{eqnarray}
\langle\sigma_{\rm eff}v\rangle&\simeq&\frac{1}{128\pi M_{\eta_1}^2}
\left(\frac{g_2^4(1+2\cos^4\theta_W)}{\cos^4\theta_W} 
+ \frac{2g_2^2g_X^2}{\cos^2\theta_W} + g_X^4\right)
\left(N_{11}+N_{22}+ 2N_{34}\right) \nonumber\\
&+&\frac{1}{32\pi M_{\eta_1}^2}
\left(\frac{g_2^4\sin^2\theta_W}{\cos^2\theta_W} + g_2^2g_X^2\right)
\left(N_{13}+N_{14}+N_{23}+
N_{24}\right) \nonumber \\
&+&\frac{1}{64\pi M_{\eta_1}^2}
\Big[\left\{\lambda_+^2+\lambda_-^2+2(\lambda_3^2+\lambda_7^2)\right\}
(N_{11}+N_{22}) \nonumber \\
&+&(\lambda_+-\lambda_-)^2(N_{33}+N_{44}+ N_{12})
+\left\{(\lambda_+ +\lambda_-)^2
+4\lambda_3^2+ 2\lambda_7^2 \right\}N_{34} \nonumber\\
&+&\left\{(\lambda_+-\lambda_3)^2+
(\lambda_- -\lambda_3)^2\right\}
(N_{13}+N_{14}+N_{23}+N_{24})\Big],
\label{cross}
\end{eqnarray}
where $g_X$ is assumed to be much smaller than $g_Y$ 
and then $X_\mu$ is sufficiently lighter than the dark matter $\eta_R$. 
$N_{ij}$ is defined by using 
$g_{\rm eff}=\sum_i \frac{n^{\rm eq}_i}{n_1^{\rm eq}}$, 
\begin{equation}
N_{ij}\equiv\frac{1}{g_{\rm eff}^2}
\frac{n_i^{\rm eq}}{n_1^{\rm eq}}\frac{n_j^{\rm eq}}
{n_1^{\rm eq}}
=\frac{1}{g_{\rm eff}^2}
\left(\frac{M_{\eta_i}M_{\eta_j}}{M_{\eta_1}^2}\right)^{\frac{3}{2}}
\exp\left[-\frac{M_{\eta_i}+M_{\eta_j}-2M_{\eta_1}}{T}\right],
\label{eqfactor}
\end{equation}
where $n_i$ is for the $\eta_i$ number density and $n_i^{\rm eq}$ is its
equilibrium value.
In order to estimate the relic abundance of $\eta_R$, we use 
the well-known analytic formula instead of solving the
Boltzmann equation numerically. The formula is given by \cite{relic},
\begin{equation}
\Omega_{\eta_1}h^2\simeq \frac{1.07\times 10^9~{\rm GeV}^{-1}}
 {J(x_F) g_\ast^{\frac{1}{2}}m_{\rm pl}},
\end{equation}
where $g_\ast$ is the relativistic degrees of freedom. 
The freeze-out temperature $T_F(\equiv \frac{M_{\eta_1}}{x_F})$ and
$J(x_F)$ are defined as
\begin{equation}
x_F=\ln\frac{0.038~ m_{\rm pl}~ g_{\rm eff}~ M_{\rm \eta_1}\langle 
\sigma_{\rm eff}v\rangle }{(g_\ast x_F)^{\frac{1}{2}}}, \qquad J(x_F)=\int^\infty_{x_F}
\frac{\langle\sigma_{\rm eff}v\rangle}{x^2}dx.
\end{equation}

\begin{figure}[t]
\begin{center}
\epsfxsize=7.5cm
\leavevmode
\epsfbox{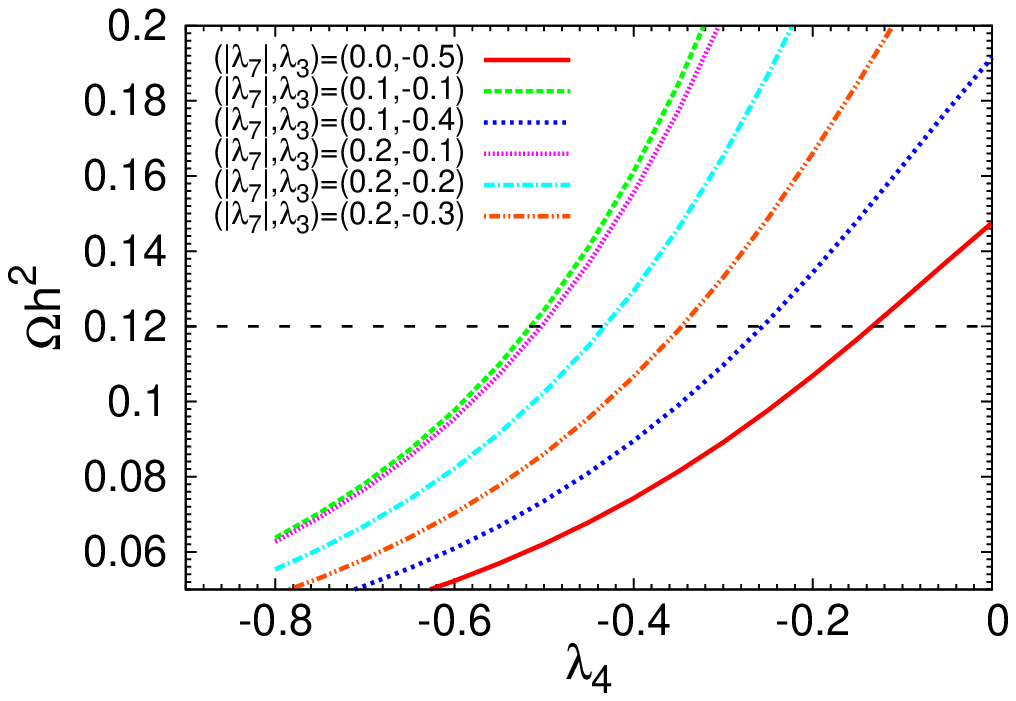}
\hspace*{5mm}
\epsfxsize=7.5cm
\leavevmode
\epsfbox{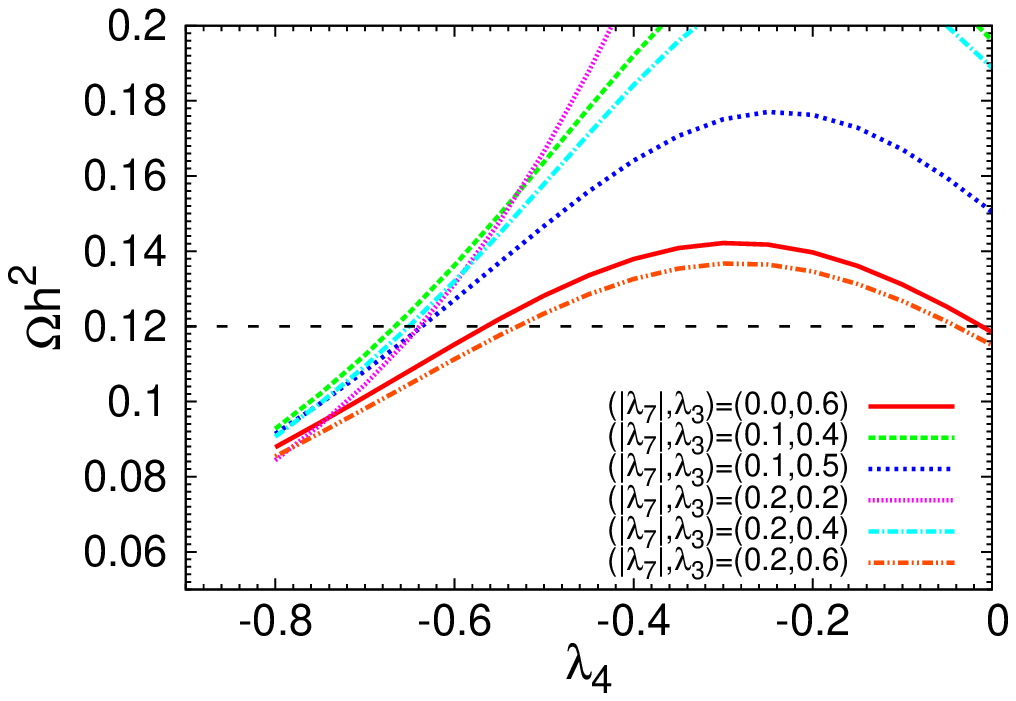}
\end{center}
\vspace*{-3mm}

{\footnotesize {\bf Fig.~3}~~Relic abundance of $\eta_R$ under 
the existence of new interactions. It is plotted as a function of 
$\lambda_4$ for typical sets of $(|\lambda_7|,\lambda_3)$. 
In the left and right panels, $\lambda_3$ is assumed to be negative 
and positive, respectively.  
A horizontal dashed line stands for the observed value 
$\Omega_{\eta_R}h^2=0.12$ \cite{planck}. 
In this plot, $g_X=0.1g_Y$ and $\lambda_5=-10^{-4}$
are assumed.}  
\end{figure}

In Fig.~3 we show the predicted relic abundance of $\eta_R$ 
when the new interactions are taken into account. 
It is plotted as a function of $\lambda_4$ 
by assuming typical values of $(|\lambda_7|,\lambda_3)$. 
To plot this figure, we assume a small value for $g_X$, such as $0.1g_Y$, 
and we fix the value 
of $m_\eta^2+\lambda_7\langle S\rangle^2$ at 1~${\rm TeV}^2$ 
for $\langle S\rangle=2$~TeV. 
Thus, the mass of $X_\mu$ is comparable to the one of the weak bosons and 
$\lambda_7$ is confined to $|\lambda_7| < 0.25$. 
The figure shows that the above cross section can explain 
the required dark matter relic abundance for a wide range of values 
of $\lambda_{3,4}$.
Since the additional (co)annihilation decay processes can generate
substantial contributions for a larger $|\lambda_7|$ in this extended model,
$|\lambda_3|$ and $|\lambda_4|$ could take much
smaller values in comparison with the values required in
the original model \cite{ks}. 
From the view point of dark matter search, however, 
the small $|\lambda_7|$ may be
promising as suggested through eq.~(\ref{direct0}). Since larger values of 
$|\lambda_{3,4}|$ are required by the relic abundance in this case,
the $\eta_R$ dark matter could be found in the Xenon1T direct search 
as discussed in \cite{ks}. On the other hand, it might be difficult 
to detecte even in the Xenon1T experiment in the case of 
a large $|\lambda_7|$. 
 
\subsection{Cosmological signal}
In this model, the main phenomenological difference from the original 
Ma model is the existence of the neutral scalar $\tilde\sigma$ and the 
neutral gauge boson $X_\mu$.\footnote{A U(1) extended model has 
been discussed in a different context \cite{mr}. }
They have no direct interaction with the 
contents of the standard model except for the one caused by 
the $\lambda_6S^\dagger S\phi^\dagger\phi$ term.
If $\tilde\sigma$ is light enough, it induces the Higgs invisible 
decay through this term as discussed already. 
Even in that case, if $\lambda_6$ satisfies the
required condition, the model is consistent with 
the present data obtained from collider experiments.
Moreover, we find no substantial constraint on the masses 
of $\tilde\sigma$ and $X_\mu$ from the study of 
the baryon number asymmetry in the previous section 
at least for the assumed value of $\langle S\rangle$. 
On the other hand, these new particles could bring about
some crucial influence to the thermal history of the Universe 
depending on their masses. 

First of all, we consider the case where $X_\mu$ is heavier than $\tilde\sigma$
and then $g_X^2>2\kappa$ is satisfied.
The new gauge boson $X_\mu$ couples only with $\tilde\sigma$, $\eta$, 
$N_i$, and $\tilde N_i$. 
Since the latter three are considered to be much heavier than $X_\mu$, 
$X_\mu$ can decay only to $\gamma\tilde\sigma$ and $\ell_\alpha\bar\ell_\beta$ 
through one-loop diagrams with $\eta$ or $N_i$ and $\tilde N_i$ in the 
internal lines. 
If we take into account that the neutrino Yukawa couplings $h_{\alpha i}$ and 
$\tilde h_{\alpha i}$ should be of $O(10^{-4})$, 
we find that the dominant contribution to the $X_\mu$ decay 
comes from the $X_\mu\rightarrow\gamma\tilde\sigma$ process. 
Its decay width can be estimated as
\begin{equation}
\Gamma_X\simeq \frac{\alpha_{\rm em}{\cal F}^2}{288(4\pi)^4} 
\frac{m_X^5}{M_{\eta_c}^4}
\left(1-\frac{M_{\tilde\sigma}^2}{m_X^2}\right)^3,
\label{crxdecay}
\end{equation}
where ${\cal F}=\lambda_7-\frac{\lambda_3\lambda_6}{2\lambda_1}$.
If we impose that $\Gamma_X ~{^>_\sim}~H$ is satisfied at the
temperature where both the freeze-out of the neutron-to-proton 
ratio and the neutrino decoupling are completed, 
${\cal F}$ is found to have a lower bound,
\begin{equation}
|{\cal F}|~{^>_\sim}~ 10^{-8}
\left(\frac{M_{\eta_c}}{1~{\rm TeV}}\right)^2
\left(\frac{300~{\rm GeV}}{m_X}\right)^{\frac{5}{2}}
\left(\frac{T}{1~{\rm MeV}}\right)
\left(1-\frac{M_{\tilde\sigma}^2}{m_X^2}\right)^{-\frac{3}{2}}.
\label{xdecay}
\end{equation}
Using the constraint on $\lambda_{1,6}$ obtained from the Higgs
sector phenomenology and the constraint on $\lambda_{3,7}$ required 
by the dark matter abundance, $|{\cal F}|$ is found to take 
a large value of $O(0.1)$.  
This suggests that $\Gamma_X>H$ could be satisfied 
at the period where the photon temperature is about 1 MeV 
even for $m_X~{^>_\sim}~O(1)$~GeV.  

Although the decay product $\tilde\sigma$ does not have direct interactions 
with the standard model contents, it can decay to them 
through loop effects. Such decay products could affect the cosmological 
thermal history depending on the time when $\Gamma_{\tilde\sigma}\simeq H$ 
is realized. 
Since the neutrino Yukawa couplings should be of $O(10^{-4})$, 
the $\tilde\sigma$ decay is dominated by a two photon final state. 
It is induced through the one-loop diagram with a charged $\eta$ 
in the internal line and the decay width can be estimated as
\begin{equation}
\Gamma_{\tilde\sigma}\simeq \frac{\alpha_{\rm em}^2{\cal F}^2}
{9216 \pi^3g_X^2}\frac{M_{\tilde\sigma}^3m_X^2}{M_{\eta_c}^4}.
\label{crsdecay}
\end{equation}
If $g_X<0.88\kappa^{\frac{3}{10}}$ is satisfied for $g_X^2 > 2\kappa$, 
$\Gamma_{\tilde\sigma}$ is larger than $\Gamma_X$. 
In such a case, $\tilde\sigma$ is expected to decay instantaneously 
after the $X_\mu$ decay yields it.
Since eq.~(\ref{xdecay}) shows that this $\tilde\sigma$ decay occurs 
at $T>1$~MeV, no cosmological effect is expected. 

In the other case, $g_X>0.88\kappa^{\frac{3}{10}}$, the decay of $\tilde\sigma$ 
occurs with a delay from its production time. 
If we use the condition $\Gamma_{\tilde\sigma} \sim H$ to make a rough 
estimation of the temperature where the $\tilde\sigma$ decay comes in the 
thermal equilibrium, we have
\begin{equation}
T\sim 54 g_\ast^{-1/4}\left(\frac{|{\cal F}|}{10^{-7}}\right)
\left(\frac{1~{\rm TeV}}{M_{\eta_c}}\right)^2
\left(\frac{m_X}{300~{\rm GeV}}\right)^{\frac{3}{2}}
\left(\frac{M_{\tilde\sigma}}{m_X}\right)^{\frac{3}{2}}
\left(\frac{\langle S\rangle}{2~{\rm TeV}}\right)
~{\rm MeV}. 
\end{equation}
From this result, we find that the $\tilde\sigma$ decay could 
occur before the neutrino decoupling as long as both $|{\cal F}|$ and $m_X$ 
take suitable values for a supposed $M_{\tilde\sigma}$. 
In this case, this decay process does not affect the neutrino 
effective number in the Universe.
For example, the light $X_\mu$ such as $m_X=O(1)$~GeV does not affect it
for $|{\cal F}|>O(10^{-4})$ as long as 
$10^{-4}m_X<M_{\tilde\sigma}<m_X$ is satisfied.

On the other hand, $\ell_\alpha\bar\ell_\beta$ could also be a dominant 
decay mode of $X_\mu$ for smaller values of $|{\cal F}|$ such as 
$|{\cal F}|~{^<_\sim}~ 10^{-7} \frac{g_X}{g_Y}
\left(\frac{\bar h}{10^{-4}}\right)^2$. 
Here, we recall that the averaged value $\bar h$ of the relevant neutrino 
Yukawa couplings $h_{\alpha i}$ is required to be of $O(10^{-4})$ 
to explain both the neutrino oscillation data 
and the baryon number asymmetry in the Universe.
Such small values of $|{\cal F}|$ could be also consistent with the 
dark matter abundance as long as $\lambda_3$ or $\lambda_4$ is of $O(1)$ and
both $|\lambda_6|$ and $|\lambda_7|$ are small enough. 
In such a case, this decay process could be in thermal equilibrium still
after the neutrino decoupling. The neutrinos produced here 
could contribute to the effective neutrino number as the non-thermal 
neutrino components. 
Although this possibility may be interesting from a cosmological view point, 
a detailed analysis is beyond the scope of this paper. 
  
Finally, we study the case where $X_\mu$ is extremely light and 
then $\tilde\sigma$ is heavier than $X_\mu$. 
In such a case, the $X_\mu$ decay could cause a cosmological problem 
generally since its decay mode is limited.
The cosmological indication could largely 
change without affecting other results of the model obtained in the 
previous part.
As an interesting example, we address the situation $m_X<2m_e$ where 
the gauge coupling $g_X$ becomes unnaturally small.\footnote{We note that
leptogenesis could occur successfully in this case as found in the 
third low of Table 2.}
There, $X_\mu$ can decay only to neutrino-antineutrino pairs
through one-loop diagrams. These non-thermally produced neutrinos 
affect the present effective neutrino number. 
Its deviation from the standard value $N_{\rm eff}=3.046$ may 
be estimated as done in \cite{dr}.

The non-thermal neutrinos make the effective neutrino number 
shift from the standard value by
\begin{equation} 
\Delta N_{\rm eff}(T)=\frac{120}{7\pi^2}\left(\frac{11}{4}\right)^{\frac{4}{3}}
\frac{\rho_\nu^{\rm nth}(T)}{T^4},
\end{equation}
where $\rho_\nu^{\rm nth}(T)$ is the energy density of non-thermally
produced neutrinos at the photon temperature $T$. 
This energy density in the co-moving volume $R^3$
evolves following the differential equation 
\begin{equation}
\frac{d(\rho_\nu^{\rm nth}R^3)}{dt}=\Gamma_X(\rho_XR^3)
-H(\rho_\nu^{\rm nth}R^3).
\end{equation}
Assuming radiation domination through this evolution, we can find the
solution
\begin{equation}
\rho_\nu^{\rm nth}R^3=m_XN_X^f\frac{1}{\sqrt{\Gamma_Xt}}\xi(t),
\end{equation}
where $\xi(t)$ is defined as 
$\xi(t)={\rm erf}(\sqrt{\Gamma_Xt})-\sqrt{\Gamma_Xt}~e^{-\Gamma_Xt}$ and  
it is reduced to $\frac{\sqrt{\pi}}{2}$ in the limit 
$\Gamma_Xt\gg 1$. $N_X^f$ stands
for the $X_\mu$ number in the co-moving volume $R^3$ at the freeze-out time
of $X_\mu$. Since it could be identified with the freeze-out time of
$\eta_R$, $X_\mu$ is relativistic there and then 
$\frac{N_X^f}{R^3}=\frac{\zeta(3)}{\pi^2}{\rm g}_XT^3$ is satisfied.
Using these, we finally obtain the deviation of the 
effective neutrino number due to the non-thermally produced neutrinos: 
\begin{eqnarray}
\Delta N_{\rm eff}&=&\frac{60\sqrt{2}\zeta(3)}{7\pi^{\frac{7}{2}}}
\left(\frac{11}{4}\right)^{\frac{4}{3}}\left(\frac{8\pi^3}{90}\right)^{\frac{1}{4}}
{\rm g}_R^{\frac{1}{4}}{\rm g}_Xm_X\sqrt{\frac{1}{\Gamma_Xm_{\rm pl}}} 
\nonumber \\
&\simeq& 0.39{\rm g}_X\left(\frac{m_X}{\rm MeV}\right)
\left(\frac{10^{-20}~{\rm MeV}}{\Gamma_X}\right)^{\frac{1}{2}},
\end{eqnarray}
where ${\rm g}_R$ is for the present degrees of 
freedom of radiation and it can be
approximated by the value of the standard model. 
This result suggests that the decay width of $X_\mu$ should be 
$\Gamma_X~{^>_\sim}~10^{-20}$~MeV for 
$X_\mu\rightarrow\nu_\alpha\bar\nu_\beta$ or 
$X_\mu\rightarrow\nu_\alpha\nu_\beta$ in order to satisfy the present 
observational results \cite{planck}. However, since the dominant
contribution comes from the latter one, which is induced through 
a one-loop diagram with the small neutrino Yukawa couplings of
$O(10^{-4})$ and also $\lambda_5$ of $O(10^{-4})$, 
the decay width is much smaller than the required value.
It means that the neutrinos produced non-thermally
through the decay of $X_\mu$ give a too large
contribution to $\Delta N_{\rm eff}$. 
Thus, the model with $m_X<2m_e$ seems to be ruled out by the 
observed effective neutrino number. 
If we introduce the kinetic term mixing for $X_\mu$
and $B_\mu$, this problem might be evaded even in such a case. 
This point is briefly discussed in the appendix. 

In the present model, the new U(1)$_X$ symmetry is assumed to be local.
Even if this symmetry is supposed to be global, the scenario 
works well in the same way. However,
the reasoning for the pairwise introduction of $N_i$ and $\tilde N_i$ 
is lost in the global U(1) case. 
The difference between them is whether the massless 
Nambu-Goldstone boson 
appears after the breaking of U(1)$_X$ symmetry or not. 
This boson behaves as dark radiation and changes 
the effective neutrino number in the Universe just in the same way
as discussed in \cite{wein}.   

\section{Conclusion}
We have considered an extension of the radiative neutrino 
mass model proposed by Ma with a low energy U(1) gauge symmetry. 
If we assume a cut-off scale of the model at $O(10^{4})$~TeV and 
the breaking of this U(1) at a rather low energy scale such as $O(1)$~TeV, 
several assumptions adopted in the original model to explain 
the neutrino masses, the dark matter abundance, and the baryon number 
asymmetry in the Universe could be closely related.
 
We have shown that the breaking of this U(1) symmetry
could give a common background for these assumptions.
Both the mass degeneracy among the right-handed neutrinos
required for the resonant decay of the lightest right-handed neutrino 
and its small neutrino Yukawa coupling required for the out-of-equilibrium 
decay could be explained by the same reasoning through this extension. 
The $Z_2$ symmetry, which forbids the tree-level neutrino mass generation 
and guarantees the dark matter stability, has the same origin as the
smallness of the quartic coupling constant between the  
Higgs doublet scalar and 
the inert doublet scalar, which is an important feature of the model 
to explain the small neutrino masses. 
It is useful to recall that these are independent assumptions 
in the original Ma model. 
We have also discussed some cosmological issues of the model
which appear to be related to this extension. The effective neutrino number
could be an interesting subject in this model. 

It is interesting that we can have an economical model which 
could explain the three big problems in the standard model 
through a simple extension of the Ma model with a low energy U(1) symmetry.
A detailed study of the model might give us a clue to the construction of a 
complete framework beyond the standard model.
We will present further results obtained from a
quantitative analysis of the related problems in the model elsewhere. 
 
\newpage 
\section*{Appendix}
In this appendix, we consider cosmological issues in the case with 
a very light $X_\mu$, where the resonant leptogenesis occurs successfully
as discussed in the text.  
In order to avoid the late time decay of $X_\mu$, we might introduce 
kinetic term mixing between the gauge fields $\hat B_\mu$ and
$\hat X_\mu$ for the gauge groups U(1)$_Y$ and
U(1)$_X$.\footnote{Kinetic term mixing of
Abelian gauge fields has been discussed in various phenomenological studies
\cite{kinet}. Recent work related to dark matter can be 
found in \cite{dm-kinet}.}
The kinetic term mixing between them may be given by
\begin{equation}
-\frac{1}{4}\hat F_{\mu\nu}\hat F^{\mu\nu}-\frac{1}{4}\hat G_{\mu\nu}
\hat G^{\mu\nu}-\frac{\sin\chi}{2}\hat F_{\mu\nu}\hat G^{\mu\nu},
\end{equation} 
where $\hat F_{\mu\nu}$ and $\hat G_{\mu\nu}$ are the field strengths of
$\hat B_\mu$ and $\hat X_\mu$, respectively.
We can diagonalize these terms by taking the canonically normalized basis
$B_\mu$ and $X_\mu$ as
\begin{equation}
\left(\begin{array}{c} \hat B_\mu \\ \hat X_\mu \end{array}\right)=
\left(\begin{array}{cc} 1 & -\tan\chi \\ 0 & \displaystyle{\frac{1}{\cos\chi}}\\
\end{array}\right)\left(\begin{array}{c} B_\mu \\ X_\mu \\ \end{array}\right).
\end{equation}
The modified U(1)$_X$ charge with this new basis is given by
\begin{equation}
Q_X=\frac{\hat Q_X}{\cos\chi}+\frac{g_Y}{g_X}Y\tan\chi,
\label{xcharge}
\end{equation}
where the U(1)$_Y$ charge and both the coupling constants $g_Y$ and 
$g_X$ are defined as the ones in the no mixing case.
This suggests that the standard model contents with $Y\not=0$ 
could couple with $X_\mu$ as long as the kinetic term mixing exists. 
As a result, the analysis of the direct search and the relic abundance 
of dark matter should be modified. 
In this case, the following new interaction should be 
added to eq.~(\ref{v-int}):   
\begin{equation}
\frac{g_X}{2}\left(\frac{1}{\cos\chi}+\frac{g_Y}{2g_X}\right)
X^\mu\left(\eta_R\partial_\mu\eta_I 
- \eta_I\partial_\mu\eta_R\right).
\label{v-int1}
\end{equation}

If the kinetic term mixing exists,
inelastic scattering of $\eta_R$ can also be brought about 
by the $X_\mu$ exchange. Since both $\eta_R$-nucleon scattering cross 
sections $\sigma_n^0(X_\mu)$ and $\sigma_n^0(Z_\mu)$, which are mediated 
by the $X_\mu$ and $Z_\mu$ exchange at zero momentum transfer, 
can be related each other as
\begin{equation}
\sigma_n^0(X_\mu)\simeq
\left(\frac{m_Z^2}{m_X^2}\tan\chi\right)^2\sigma_n^0(Z_\mu),
\end{equation}
the present experimental results require that the kinetic term mixing should
satisfy 
\begin{equation}
\tan\chi~{^<_\sim}~\frac{m_X^2}{m_Z^2}.
\label{acon}
\end{equation} 
This shows that the kinetic term mixing should be sufficiently small for
$m_X~{^<_\sim}~ O(1)$~GeV.
New non-negligible (co)annihilation modes of $\eta_R$ to the standard 
model contents could also
appear, depending on the magnitude of the kinetic term mixing $\sin\chi$.
However, the constraint (\ref{acon}) suggests that the $\eta_R$ 
relic abundance could not be affected by the process mediated through the 
$X_\mu$ exchange. 
In the study of the $\eta_R$ relic abundance, even if we introduce 
the kinetic term mixing, we can neglect the effect of it 
as long as the condition (\ref{acon}) is satisfied. 
Thus, the results obtained in this paper do not change.

As another interesting phenomenon caused by the kinetic term mixing, 
we consider the $X_\mu$ direct decay to the lighter fermions in the 
standard model through tree diagrams. 
Its decay width could be estimated as 
\begin{equation}
\Gamma_X(f\bar f)=\sum_f\frac{g_Y^2}{16\pi m_X}
\left(\frac{Y_f}{2}\right)^2\tan^2\chi. 
\end{equation}
If we impose that $\Gamma_X~{^>_\sim}~H$ is satisfied before the
neutrino decoupling, we find that the kinetic term mixing should satisfy 
\begin{equation}
\tan\chi~{^>_\sim}~10^{-11}\left(\frac{1~{\rm GeV}}{m_X}\right)^{\frac{1}{2}}
\left(\frac{T}{1~{\rm MeV}}\right).
\label{bcon}
\end{equation}
This shows that a sufficiently small kinetic term mixing is enough to 
bring about the $X_\mu$ decay to the standard model fermions 
before the neutrino decoupling.
As long as the very small kinetic term mixing exists,
the model can overcome the cosmological difficulty for the effective neutrino
number in both cases with $m_X>M_{\tilde\sigma}$ and $m_X<M_{\tilde\sigma}$.
Especially, if the kinetic term mixing takes a suitable value in the case
$m_X< 1$~MeV, the deviation of the effective neutrino number 
$N_{\rm eff}=3.62\pm 0.25$, which is suggested through the
combined analysis of the data from Planck and the $H_0$ measurement from 
the Hubble Space Telescope \cite{planck}, might be explained.

\section*{Acknowledgements} 
S.~K. is supported by Grant-in-Aid for JSPS fellows (26$\cdot$5862).
D.~S. is supported by JSPS Grant-in-Aid for Scientific
Research (C) (Grant No. 24540263) and MEXT Grant-in-Aid 
for Scientific Research on Innovative Areas (Grant No. 26104009).

\newpage
\bibliographystyle{unsrt}

\begin{thebibliography}{99}
\bibitem{lhc}The ATLAS Collaboration, G.Aad, {\it et al.}, 
Phys. Lett. {\bf B716} (2012) 1; 
The CMS Collaboration, S.Chatrchyan, {\it et al.}, 
Phys. Lett. {\bf B716} (2012) 30.

\bibitem{nexp}Super-Kamiokande Collaboration, Y.~Fukuda, {\it et al.},
	Phys. Rev. Lett. {\bf 81} (1998) 1562; 
       SNO Collaboration, Q.~R~.Ahmad, {\it et al.},
	Phys. Rev. Lett. {\bf 89} (2002) 011301;
         KamLAND Collaboration, K.~Eguchi, {\it et al.}, 
       Phys. Rev. Lett. {\bf 90} (2003)
	021802; 
       K2K Collaboration, M.~H.~Ahn, {\it et al.},
	Phys. Rev. Lett. {\bf 90} (2003) 041801.    

\bibitem{t13}T2K Collaboration, K.~Abe, {\it et al.},
Phys. Rev. Lett. {\bf 107} (2011) 041801; 
Double Chooz Collaboration, Y.~Abe, {\it et al.},
	Phys. Rev. Lett. {\bf 108} (2012) 131801;
RENO Collaboration, J.~K.~Ahn, {\it et al.}, 
Phys. Rev. Lett. {\bf 108} (2012) 191802;
The Daya Bay Collaboration, F.~E.~An, {\it et al.},
	Phys. Rev. Lett. {\bf 108} (2012) 171803. 

\bibitem{uobs}WMAP Collaboration, D.~N.~Spergel, {\it et al.},
	Astrophys. J. Suppl. {\bf 148} (2003) 175; 
SDSS Collaboration, M.~Tegmark, {\it et al.}, 
Phys. Rev. {\bf D69} (2004) 103501.

\bibitem{planck}Planck Collaboration, P.~A.~R.~Ade, {\it et al.}, 
Astron. Astrophys. {\bf 571} (2014) A16; 
Planck Collaboration, P.~A.~R.~Ade, {\it et al.}, 
arXiv:1502.01589 [astro-ph.CO].

\bibitem{basym}A.~Riotto and M.~Trodden,
	Ann. Rev. Nucl. Part. Sci. {\bf 49} (1999) 35; 
W.~Bernreuther, Lect. Notes
	Phys. {\bf 591} (2002) 237; M.~Dine and A.~Kusenko,
	Rev. Mod. Phys. {\bf 76} (2003) 1.

\bibitem{ma}E.~Ma, Phys. Rev. {\bf D73} (2006) 077301.

\bibitem{ndm}J.~Kubo, E.~Ma and D.~Suematsu, Phys. Lett. {\bf B642} (2006) 18;
D.~Aristizabal Sierra, J.~Kubo, D.~Restrepo,
D.~Suematsu and O.~Zapata, Phys. Rev. {\bf D79} (2009) 013011;
D.~Suematsu, T.~Toma and T.~Yoshida, Phys. Rev. {\bf D79} (2009) 093004; 
D.~Suematsu, T.~Toma and T.~Yoshida, Phys. Rev. {\bf D82} (2010) 013012. 

\bibitem{u1}J.~Kubo and D.~Suematsu, Phys. Lett. {\bf B643} (2006) 336;
D.~Suematsu, Eur. Phys. J. {\bf C56} (2008) 379.

\bibitem{susyndm}H.~Fukuoka, J.~Kubo and D.~Suematsu, Phys. Lett. {\bf B678}
	(2009) 401; D.~Suematsu and T.~Toma, Nucl. Phys. {\bf B847}
	(2011) 567; H.~Fukuoka, D.~Suematsu and T.~Toma, 
JCAP {\bf 1107} (2011) 001.

\bibitem{ndm1}H.~Higashi, T.~Ishima and D.~Suematsu, 
Int. J. Mod. Phys. {\bf A26} (2011) 995; 
D.~Suematsu, Eur. Phys. J {\bf C72} (2012) 72; 
D.~Suematsu, Phys. Rev. {\bf D85} (2012) 073008.

\bibitem{ks}S.~Kashiwase and D.~Suematsu, Phys. Rev. {\bf D86} (2012)
	053001; S.~Kashiwase and D.~Suematsu, Eur Phys. J. {\bf C73} (2013)
	2484. 

\bibitem{infl}R.~H.~S.~Buhdi, S.Kashiwase and D.~Suematsu, Phys. Rev. {\bf D90} 
(2014) 113013; R.~H.~S.~Buhdi, S.Kashiwase and D.~Suematsu, JCAP 
{\bf 1509} (2015)
039; S.~Kashiwase and D.~Suematsu, Phys. Lett. {\bf B749} (2015) 603;
R.~H.~S.~Buhdi, S.Kashiwase and D.~Suematsu, Phys. Rev. {\bf D93} (2016) 013022.

\bibitem{res}M.~Flanz, E.~A.~Paschos and U.~Sarkar, 
Phys. Lett. {\bf B345} (1995) 248; L.~Covi, E.~Roulet and F.~Vissani,
	Phys. Lett. {\bf B384} (1996) 169;
E.~Akhmedov, M.~Frigerio and A. Yu Smirnov, JHEP {\bf 0309} (2003) 021;
C.~H.~Albright and S.~M.~Barr, Phys. Rev. {\bf D69} (2004) 073010. 

\bibitem{res-f1}A.~Pilaftsis, Phys. ReV. {\bf D56} (1997) 5431; 
T.~Hambye,J.~March-Russell and S.~W.~West, JHEP {\bf 0407} (2004) 070; 
A.~Pilaftsis and T.~E.~J.~Underwood, Phys. Rev. {\bf D72} (2005) 113001.

\bibitem{res-f2}A.~Pilaftsis and T.~E.~J.~Underwood, 
Nucl. Phys. {\bf B692} (2004) 303.

\bibitem{pdg}J.~Beringer {\it et al.} (Particle Data Group),
	Phys. Rev. {\bf D86} (2012) 010001.

\bibitem{res2}P.~S.~Bhupal Dev, P.~Millington, A.~Pilatfsis and D.~Teresi,
Nucl. Phys. {\bf B886} (2014) 569.

\bibitem{idm}R.~Barbieri, L.~J.~Hall and V.~S.~Rychkov, Phys. Rev. {\bf
	D74} (2006) 015007; M.~Cirelli, N.~Fornengo and A.~Strumia,
	Nucl. Phys. {\bf B753} (2006) 178; L.~L.~Honorez, E.~Nezri,
	J.~F.~Oliver and M.~H.~G.~Tytgat, JCAP {\bf 0702} (2007) 028; 
Q.-H.~Cao, E.~Ma and G.~Rajasekaran, Phys. Rev. {\bf D76} (2007) 095011;
S.~Andreas, M.~H.~G.~Tytgat and Q.~Swillens, JCAP {\bf 0904} (2009) 004; 
E.~Nezri, M.~H.~G.~Tytgat and G.~Vertongen, JCAP {\bf 0904} (2009) 014;
L.~L.~Honorez and C.~E.~Yaguna, JCAP {\bf 1101} (2011) 002; 
M.~Gustafsson, S.~Rydbeck,
	L.~L.~Honorez, and E.~Lundstr\"{o}m, Phys. Rev. 
{\bf D86} (2012) 075019.

\bibitem{idm1}T.~Hambye, F.-S.~Ling, L.~L.~Honorez and J.~Rocher, JHEP
	{\bf 0907} (2009) 090.  

\bibitem{stab}S.~Nie and M.~Sher, Phys. Lett. {\bf B449} (1999) 89;
S.~Kanemura, T.~Kasai and Y.~Okada, Phys. Lett. {\bf B471} (1999) 182.

\bibitem{hwidth}P.~P.~Giardino, K.~Kannike, I.~Masina, M.~Raidal and
	A.~Strumia, JHEP {\bf 1405} (2014) 046.

\bibitem{wein}S.~Weinberg, Phys. Rev. Lett. {\bf 110} (2013) 241301.

\bibitem{pseudoD}For example, M.~B.~Gavela, T.~Hambye, D.~Hernandez and
	P.~Hernandez, JHEP {\bf 0909} (2009) 038; D.~A.~Sierra, A.~Degee
	and J.~F.~Kamenik, JHEP {\bf 1207} (2012) 135.

\bibitem{inel}D.~T.~Smith and N.Weiner, Phys.Rev.{\bf D72} (2005) 063509; 
S.~Chang, G.~D.~Kribs D.~T.~Smith and N.~Weiner, Phys. Rev. {\bf D79}
	(2009) 043513

\bibitem{l5}Y.~Cui, D.~E.~Marrissey, D.~Poland and L.~Randall, JHEP {\bf
	0905} (2009) 076;
C.~Arina, F.-S.~Ling and M.~H.~G.~Tytgat, JCAP {\bf 0910}
	(2009) 018.

\bibitem{inelvel}D.~Smith and N.~Weiner, Phys. Rev. {\bf D64} (2001)
	043502.

\bibitem{direct1}CDMS Collaboration, Z.~Ahmed, {\it et al.},
	Phys. Rev. Lett. {\bf 102} (2009) 011301; XENON100
	Collaboration, E.~Aprile, {\it et al.}, Phys. Rev. Lett. {\bf
	105} (2010) 131302.

\bibitem{lux}LUX Collaboration, D.~S.~Akerib, {\it et al.},
Phys. Rev. Lett. {\bf 112} (2014) 091303.

\bibitem{direct2}G.~Angloher {\it et al.}, Astropart. Phys. {\bf 31}
	(2009) 270; V.~N.~Lebedenko {\it et al.}, Phys. Rev. {\bf D80}
	(2009) 052010.

\bibitem{relic}K.~Griest and D.~Seckel, Phys. Rev. {\bf D43} (1991)
	3191; P.~Gondolo and G.~Gelmini, Nucl. Phys. {\bf B360} (1991) 145.

\bibitem{mr}E.~Ma, I.~Picek and B.~Radovcic, 
Phys. Lett. {\bf B726} (2013) 744.

\bibitem{dr}P.~Di Bari, S.~F.~King and A.~Merle, 
Phys. Lett. {\bf B724} (2013) 77.

\bibitem{kinet}B.~Holdom, Phys. Lett. {\bf B166} (1986) 196; T.~Matsuoka
	and D.~Suematsu, Prog. Theor. Phys. {\bf 76} (1986) 901;
	K.~R.~Dienes, C.~F.~Kolda and J.~March-Russell, Nucl. Phys. {\bf
	B492} (1997) 104; B.Holdom, Phys. Lett. {\bf B259} (1991) 329;
K.~S.~Babu, C.~F.~Kolda and J.~March-Russell, Phys. Rev. {\bf D57}
	(1998) 6788; D.~Suematsu, Phys. Rev. {\bf D59} (1999) 055017;
D.~Suematsu, JHEP {\bf 0611} (2006) 029. 

\bibitem{dm-kinet}R.~Foot and R.~R.~Volkas, Phys. Rev. {\bf D52} (1995)
	6595; C.~Boehm and P.~Fayet, Nucl. Phys. {\bf B683}
	(2004) 219; C.~Boehm, P.~Fayet and J.~Silk, Phys. Rev. {\bf D69}
	(2004) 101302;  M.~Pospelov, A.~Ritz and M.~B.~Voloshin, 
        Phys. Lett. {\bf B662} (2008) 53; 
         J.~L.~Feng and J.~Kumar, Phys. Rev. Lett. {\bf 101}
	(2008) 231301; D.~Hooper and K.~M.~Zurek, Phys. Rev. {\bf D77}
	(2008) 087302; N.~Arkani-Hamed
	and N.~Weiner, JHEP {\bf 0812} (2008) 104; 
        N.~Arkani-Hamed, D.~P.~Finkbeiner, T.~R.~Slatyer and
	N.~Weiner, Phys. Rev. {\bf D79} (2009) 015014.
	 
\end{thebibliography}

\end{document}